\documentclass[]{aastex631}
\usepackage{amsmath}	%
\usepackage{bm}        %
\usepackage{placeins}
\usepackage{newtxtext,newtxmath}
\usepackage{esvect}
\usepackage{xcolor}

\newcommand{\bigdot}[1]{\overset{\bm .}{#1}}

\newcommand\degree{$^{\circ}$}

\shorttitle{}
\shortauthors{Chen et al.}
\graphicspath{{./}{}}
\begin{document}

\title{Transient Blurring of the Scintillation Arc of Pulsar B1737+13}

\author{Yen-Hua Chen}
\affiliation{Institute of Astronomy and Astrophysics, Academia Sinica, Taipei 10617, Taiwan}
\affiliation{Department of Mechanical Science \& Engineering, University of Illinois Urbana-Champaign, Urbana, IL 61801, USA}
\email{yenhuac2@illinois.edu}
\author{Sammy Siegel}
\affiliation{Department of Astronomy, Boston University, Boston, US}
\affiliation{Department of Physics and Astronomy, Oberlin College, Oberlin, OH 44074, USA}
\author{Daniel Baker}
\affiliation{Institute of Astronomy and Astrophysics, Academia Sinica, Taipei 10617, Taiwan}
\author{Ue-Li Pen}
\affiliation{Institute of Astronomy and Astrophysics, Academia Sinica, Taipei 10617, Taiwan}
\affiliation{Department of Physics, University of Toronto, Toronto, Canada}
\affiliation{Canadian Institute for Theoretical Astrophysics, University of Toronto, Toronto, Canada}
\author{Dan Stinebring}
\affiliation{Department of Physics and Astronomy, Oberlin College, Oberlin, OH 44074, USA}

\begin{abstract}
For many pulsars, the scattering structures responsible for scintillation are typically dominated by a single, thin screen along the line of sight, which persists for years or decades. In recent years, an increasing number of doubly-lensed events have been observed, where a secondary lens crosses the line of sight. This causes additional or distorted scintillation arcs over time scales ranging from days to months. In this work we report such a transient event for pulsar B1737+13 and propose a possible lensing geometry including the distance to both lenses, and the orientation of the main screen. Using phase retrieval techniques to separate the two lenses in the wavefield, we report a curvature and rate of motion of features associated with the secondary lens as it passed through the line of sight. By fitting the annual variation of the curvature, we report a possible distance and orientation for the main screen. The distance of the secondary lens is found by mapping the secondary feature onto the sky and tracking its position over time for different distances. We validate this method using B0834+06, for which the screen solutions are known through VLBI, and successfully recover the correct solution for the secondary feature. 
With the identified lensing geometry, we are able to estimate the size of the secondary lens, 1 -- 3~au.
Although this an appropriate size for a structure that could cause an extreme scattering event, we do not have conclusive evidence for or against that possibility.

\end{abstract}


\section{Introduction} \label{sec:intro}
Pulsars are excellent sources of spatially coherent radio emission, which allows them to exhibit interference phenomena when scattered by the interstellar medium (ISM). 
Multi-path scattering of this coherent radiation by inhomogeneities in the ISM, coupled with relative motion of the pulsar (often dominant), the Earth, and ISM structures produces a modulated radio spectrum that changes with time.
Instead of observing pulsar images directly, we measure intensity fluctuations (scintillation) on a time-frequency plane and infer information about the imaging from that scintillation pattern.
The resolution of this method depends on the angular size of the scatter-broadened image and  generally exceeds
the diffraction limit of Earth-based interferometers \citep[e.g.][]{Brisken2010}.
Hence, detailed studies of pulsar scintillation provide valuable insights into the structure of ISM over a wide range of spatial scales. 

\begin{figure}
    \includegraphics[width=\columnwidth]{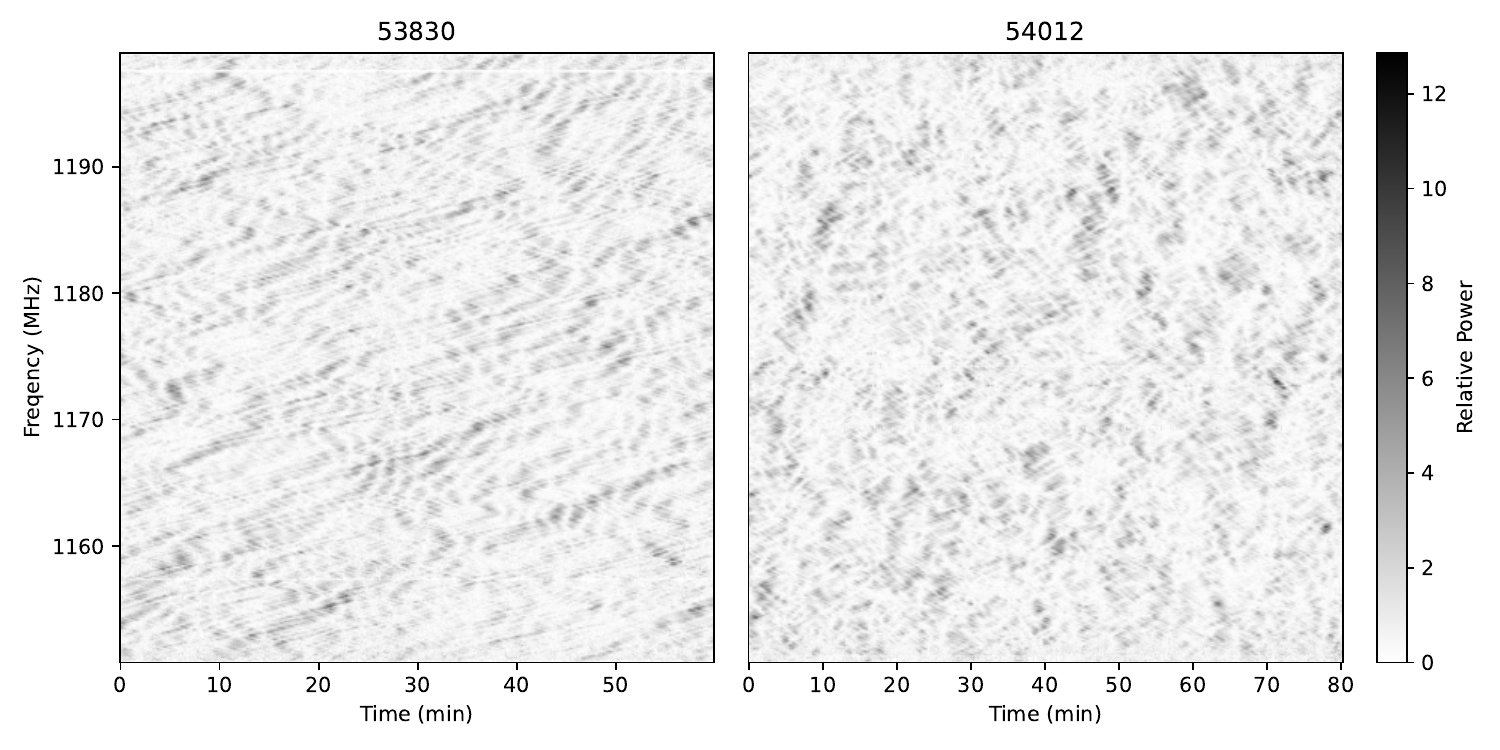}
    \caption{Dynamic spectrum of MJD~53830 (left) and MJD~54012 (right) in the 1175 MHz band. The time-dependent variation patterns are caused by the scintillation of ISM. The scale of patterns is larger than time resolution, and we take this advantage in the signal correlation in later analysis. }
    \label{fig:53830_54012_DS}
\end{figure}

Classical pulsar scintillation studies 
concentrated on characterizing the typical frequency and time scale of the 
scintillation features and using them to infer general properties of the ISM \citep{Lyne1984,ric90,Cordes1998,Bogdanov2002,Rickett2014}.
In general, this was done by using either the 1D or 2D auto-correlation functions of the 
dynamic spectra, the time-dependent variation of the pulsed radio spectrum of the pulsar. Figure~\ref{fig:53830_54012_DS} provides an example of dynamic spectra in B1737+13. With the discovery of scintillation arcs \citep{smc+01}, indicative of scattering
occurring in localized regions or ``screens'' along the line of sight,
emphasis has focused on analyzing scintillation in the so-called conjugate spectrum, which is the Fourier transform of the dynamic spectrum. (For a more detailed review of the various spectra and their transformations used in scintillation, see Table 1 in \citet{baker-2023}). In this work, we will use lens to refer to any refractive structure in the ISM that produces one or more images of the pulsar, while screen will be used specifically to refer to extended lenses at a fixed distance that form multiple images over a range of angles.

Mapping into the conjugate spectrum offers several advantages over the dynamic spectra for identifying individual screens and features on the sky. While every point in the dynamic spectrum is the result of interference between all scattered images, in the simple case of a single 1D screen, each point in the conjugate spectrum provides information about the interference between a single pair of images. In addition, pairs that share an image are visually grouped in the conjugate spectrum space. For the simple one-dimensional case, these groups lie along inverted arclets whose apexes, corresponding to interference between the shared image and the line of sight, all lie along a single forward parabola of the same curvature, $\eta$. Since the curvature for distinct screens will be typically be different, this allows us to identify each screen in the conjugate spectrum. Several examples of this phenomena are explored in \cite{Ocker2023}. As such, the analysis in this paper focuses on the conjugate spectra. However, for ease of visualization, we will generally display its squared magnitude, known as the secondary spectra.

Consider a simple one-dimensional line of images with a particular image located at position  $\vv{\theta}$ relative to the origin. The observed parabolic structure of the main arc arises naturally in such a picture.
The Doppler shift $f_{\rm{D}}$ from the relative motion of Earth, ISM, and pulsar and the time delay $\tau$ from the additional path length are given by

\begin{equation}
    \tau = \frac{ d_{\rm eff}\; \theta^{2}}{2c} 
    \label{eq:tau}
\end{equation}
\begin{equation}
    f_D = \frac{-\vv{V}_{\rm{eff}}\cdot \vv{\theta}}{\lambda}
    \label{eq:fd}
\end{equation}

\noindent where $\lambda$ is the observing wavelength, $d_{\rm{eff}} = d_{\rm psr}(1-s)/s$, and $\vv{V_{\rm{eff}}} = \vv{V_\earth} -  \vv{V_{\rm scr}}/s +  \vv{V_{\rm psr}}(1-s)/s$ are the effective distance to the screen and effective velocity along the line of images, respectively; here, $0 < s < 1$ is the  distance from the pulsar to the screen as a fraction of the total Earth - pulsar distance. For more detailed discussion of these relationships and scintillation arc structures, see \cite{wmsz04,crsc06} and \cite{swm+21}. 

Equations \ref{eq:tau} and \ref{eq:fd} yield an outer boundary to the scintillation arc
\begin{equation}
    \tau = \eta f_{\rm{D}}^2
    \label{eq:quadratic eqn}
\end{equation}
with the curvature, $\eta$, given by
\begin{equation}
    \eta = \frac{d_{\text{eff}} \lambda^2}{2c V_{\rm{eff}}^2}.
    \label{eq:curvature}
\end{equation}
The variation of this curvature over the course of Earth's orbit, or the pulsar's orbit if it is in a binary system, encodes information about the distance to and orientation of the screen \citep[e.g.][]{rcb+20}. As lenses that form only a single image are just a special case of the one-dimensional lens, curvature measurements can also be useful for studying their properties. 

By combining measurements of the scintillation arc with very long baseline interferometry (VLBI), \cite{Brisken2010} produced the first high-resolution images of a pulsar scintillation screen. Their work not only localized the screen along the line of sight, but also imaged the screen and showed, for the first time, that the lensing truly did occur along a single nearly straight line. In addition to the main arc, they also observed a secondary parabolic structure offset from the main arc at a delay of approximately $1~\rm{ms}$ that we refer to as the ``millisecond feature.'' They found this feature to also be offset on the sky and at a different distance from the main arc. \cite{Zhu2023} discuss this secondary feature more deeply and follow its evolution over a series of observations over a few weeks surrounding the initial observation. With VLBI data, all the information needed for modeling, including distances, velocities, and orientations are known. Based on those results, they were able to set up a double-lensing model to simulate the scintillation. In their simulation, the motion of the secondary feature in the wavefield matched the observation quite well, which strongly supported this scenario.
In addition to modeling the millisecond feature, \cite{Zhu2023} also discuss the possibility of whether the structure responsible for the feature could also cause extreme scattering events (ESEs). First reported by \cite{Fiedler1987} through the observation of extragalactic radio source, ESEs are an unusual change in the flux of steady radio sources. ESEs are typically identified by a period of constant reduced intensity, sandwiched between two brightness peaks. Due to their strong frequency dependence, they are unlike intrinsic variations of these sources and are generally
attributed to lensing within the ISM \citep{Fiedler1987}. 

One potential explanation for scattering to form a line of images is the corrugated sheet picture proposed by \cite{Pen2014} and later expanded upon by \cite{Simard2018}. In this scenario, lensing occurs at the folds along a thin plasma sheet produced by magnetic reconnection between two regions of different magnetic fields.
These events seem to arise naturally in the corrugated sheet model at cusps as described by \cite{Jow2023}. In their model, at the ends of folds in a corrugated sheet they merge and flatten out at cusps. This creates a longer region of the sheet aligned parallel to the line of sight than in typical scintillation, allowing for the large bending angles needed to form ESEs without the need for extreme changes in the electron density. The connection between scintillation and ESEs is further strengthened by \cite{Zhu2023}, who demonstrated that the size and occurrence rate of these features is consistent with the duration and rate of ESEs. With this picture in mind, it is useful to search through archival data in order to locate other such events. In this paper, we report on one such event seen in B1737+13.
During this event, we see that the secondary spectra make a transition from clear inverted arclets to a fuzzy period and back again over the course of several weeks.

The rest of the paper is organized as follows.
Details about the observations are given in \S\ref{sec:obs}.
Our analysis of the data is presented in \S\ref{sec:analysis} and consists of the following steps.
First, in \S\ref{sec:thth-meas} we use the $\theta-\theta$ method to measure curvatures throughout the event and identify features lying along a different parabola than the main arc. 
In \S\ref{sec:Annual Fitting}, we propose a possible distance for the main screen by annual-fitting. 
Then, in \S\ref{sec:secondary-feature} and \S\ref{sec:secondary-curvature} we make a measurement of the curvature of the secondary arc and track the movement of a collection of images along it. 
For the secondary lens, we localize it by mapping the observed images onto the sky for different potential lens distances and minimize their motion across the sky over time in \S\ref{sec:image_ont_the_sky}.
In \S\ref{sec:Possible Physical Picture} we build a simulation to support the solutions we found in the previous sections.
We then use the lensing geometry to determine the size of the secondary lens in \S\ref{sec:Size of the secondary screen}. 
Based on the evidence developed for the unusual features of the secondary lens, we then  make a connection of this event to Extreme Scattering Events in \S\ref{sec:ESE}.
Finally, we summarize our findings in \S\ref{sec:Conclusion}.
In Appendix~\ref{sec:Appendix_A}, we introduce the concept of interaction arcs to explain the ``fuzziness'' seen during a double lensing event like the one reported here. 
In Appendix~\ref{sec:Appendix_B}, we apply the displacement test described in \S~\ref{sec:Possible Physical Picture} to data of pulsar B0834+6 to show the viability of the method when compared to VLBI results.

\section{Observations}
\label{sec:obs}

We use a series of observations of pulsar B1737+13 that were first reported in \cite{2008ApJ...674L..37H} and taken over nearly 37 weeks with the Arecibo telescope from April 5 to December 31 in 2006 (MJD~53830 to 54100). The observations consist of four bands with center frequencies at 1175, 1380, 1425, and 1470 MHz, each having a bandwidth of 50 MHz. The observational setup used 2048 channels per band, providing a frequency resolution of 24.4 kHz, with subintegration times of 10~s. The proper motion of B1737+13 is $[-22,-20] \pm 2 \; \rm{mas}\, \rm{yr}^{-1}$ \citep{Brisken_2003} , and the distance obtained from its dispersion measure (DM) is 4.2~kpc. 

\begin{figure}
    \includegraphics[width=\columnwidth]{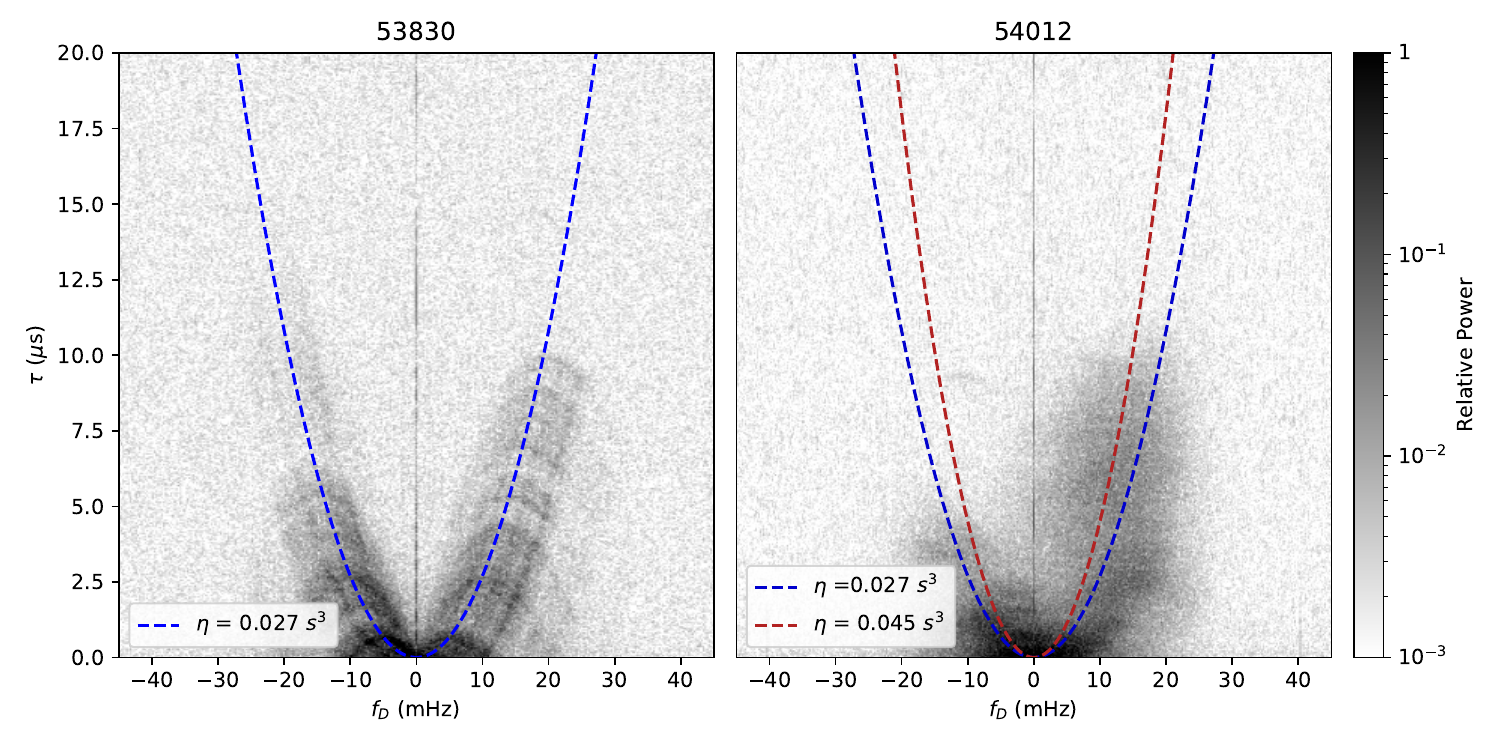}
    \caption{Conjugate spectrum of MJD~53830 (left) and MJD~54012 (right) in the 1175 MHz band. The left panel (MJD~53830) shows the main parabolic arc with inverted arclets, representing a classic single-screen scintillation pattern. In contrast, the right panel (MJD~54012) reveals a significantly fuzzier arc, suggesting the presence of a secondary lens influencing the scintillation pattern. The blue dashed line indicates the primary screen curvature, $\eta = 0.027~{\rm s}^3$, while the red dashed line, with $\eta = 0.045~{\rm s}^3$, is approximated curvature of secondary lens. The color bar on the right displays the relative power, with darker regions indicating stronger intensity. The change in curvature and arc clarity between these two epochs indicates that a transient event happened in B1737+13.}
    \label{SS53830_54012}
\end{figure}

From MJD~53830 to approximately 53978, the secondary spectra have fairly typical arc and inverted arclet structures, as seen on the left hand side of Figure~\ref{SS53830_54012}. However, around MJD~53978, a secondary feature at a higher curvature 
enters the secondary spectra. As this features moves towards the origin, the main arc becomes less sharp, as seen in the right hand panel of Figure~\ref{SS53830_54012} for MJD~54012. Eventually, this feature exits the secondary spectra, and the main arc returns to normal around MJD~54050.

\section{Data Analysis}
\label{sec:analysis}
The first step of the analysis is measuring the variation in the curvature of the lenses, allowing the determination of their locations and further modeling.

\subsection{$\theta - \theta$ Measurement}
\label{sec:thth-meas}

\begin{figure}
    \centering
    \includegraphics[width=0.5\columnwidth, height=0.5\columnwidth]{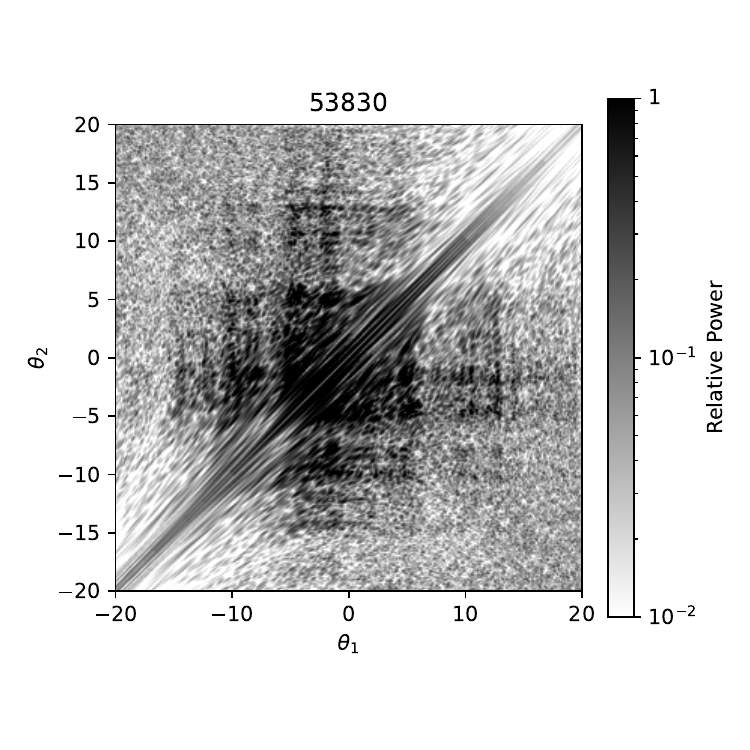}
    \caption{The $\theta_1-\theta_2$ map for MJD~53830 shows clear linear pattern, confirming that a curvature of 0.027~${\rm s}^3$ provides the best fit for this scintillation arc.}
    \label{fig:thth53830}
\end{figure}

\begin{figure}
    \centering
    \includegraphics[width=\columnwidth]{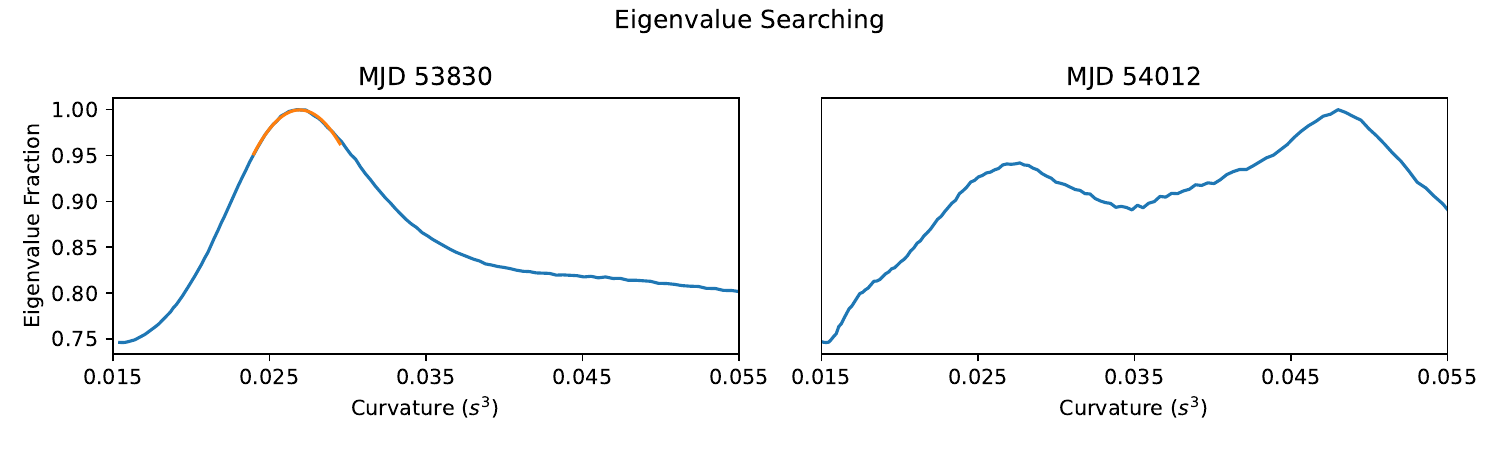} 
    \caption{Value of the largest eigenvalue of the $\theta-\theta$ matrix for different curvature for MJD~53830 (left) and MJD~54012 (right) scaled relative to the peak. The left plot (MJD~53830) shows a successful example of eigenvalue searching, with a clear peak representing the best-fitting curvature, which has the largest eigenvalue among the tested curvatures. In contrast, the right plot (MJD~54012) displays double peaks due to the presence of multiple lens in the data, indicating that the influence of the secondary lens is comparable to that of the main screen.}
    \label{fig:eigenvalue_53830_54012}
\end{figure}

\begin{figure}
    \centering
    \includegraphics[width=0.8\columnwidth]{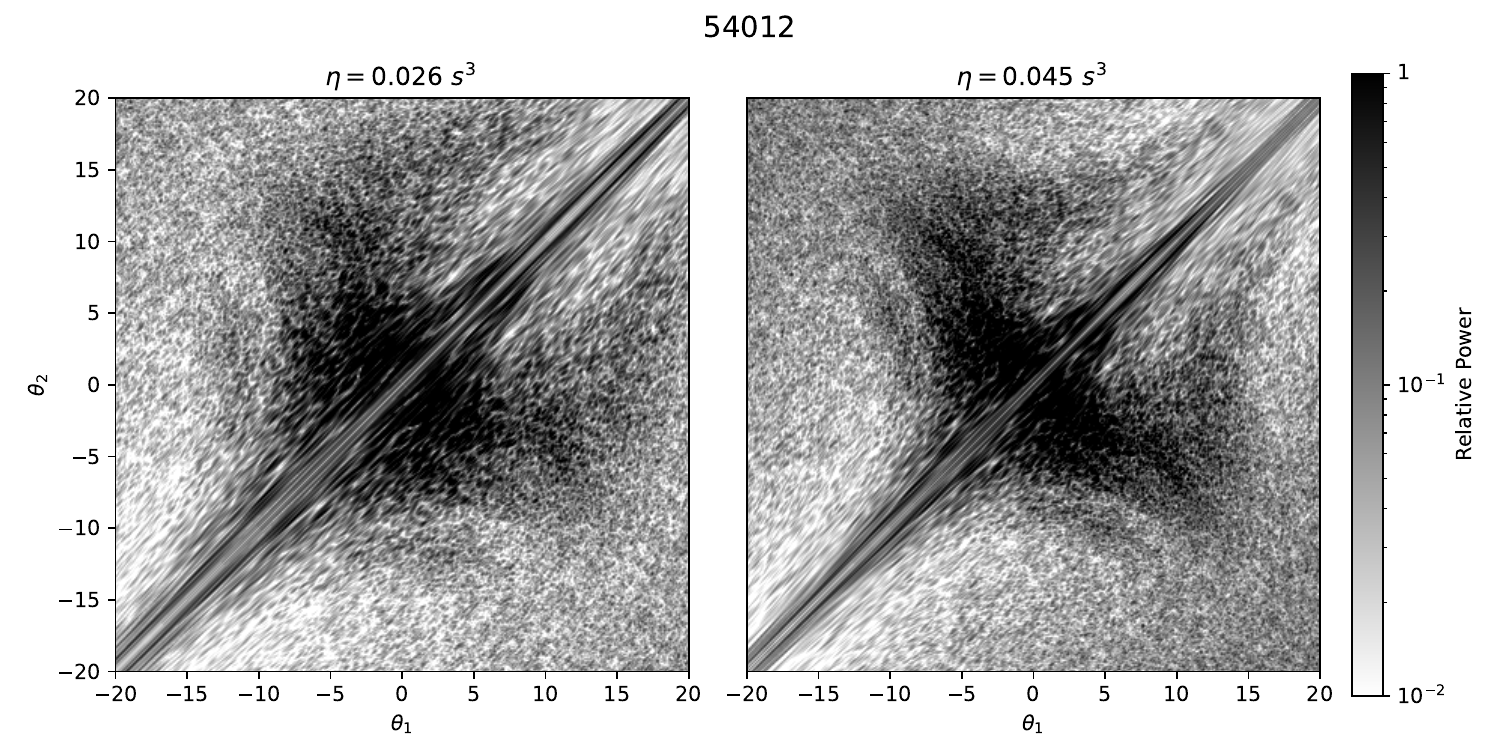}
    \caption{$\theta-\theta$ maps for MJD~54012 using the persistent arc curvature of $0.026~\rm{s}^3$ (left) and the transient feature curvature of $0.045~\rm{s}^3$ (right). In the left panel, the pattern is more rectangular for small $\theta_1$ and $\theta_2$, but is skewed for the more distant features in the upper right. In contrast, the pattern in right panel is skewed near the origin and more rectangular for the distant features.}
    \label{fig:thth54012_2745}
\end{figure}

The curvature, $\eta$, is the coefficient of the quadratic equation (\ref{eq:quadratic eqn}), which describes the parabolic scintillation arc in the conjugate spectrum. We measure this curvature using the $\theta-\theta$ transform, as proposed by \cite{TimSprenger2021} and further developed by \cite{Baker2022}. In $\theta-\theta$ space, the main arc and the inverted arclets become straight lines with their orientation set by the curvature used in the transformation. For the optimal curvature, all the curves in the conjugate spectrum are transformed into lines parallel to the axes as seen in Figure~\ref{fig:thth53830}, and the dominant eigenvalue is maximized. For any other curvature, the pattern will be skewed along the diagonals
and power will be shared among several eigenvalues. For observations such as MJD~53830 where the arclets are well defined, a search of the largest eigenvalue as a function of curvature shows a single clear peak as in the left panel of Figure~\ref{fig:eigenvalue_53830_54012}. However, if there are multiple features with different curvatures, such as at MJD~54012, there will be multiple peaks with more substructure in the eigenvalue search as shown in the right panel of Figure~\ref{fig:eigenvalue_53830_54012} and skewed structure in the $\theta-\theta$ plots as shown in Figure~\ref{fig:thth54012_2745}.

\begin{figure}[htp]
    \begin{minipage}[t]{0.545\textwidth}
    \centering    \includegraphics[width=\textwidth]{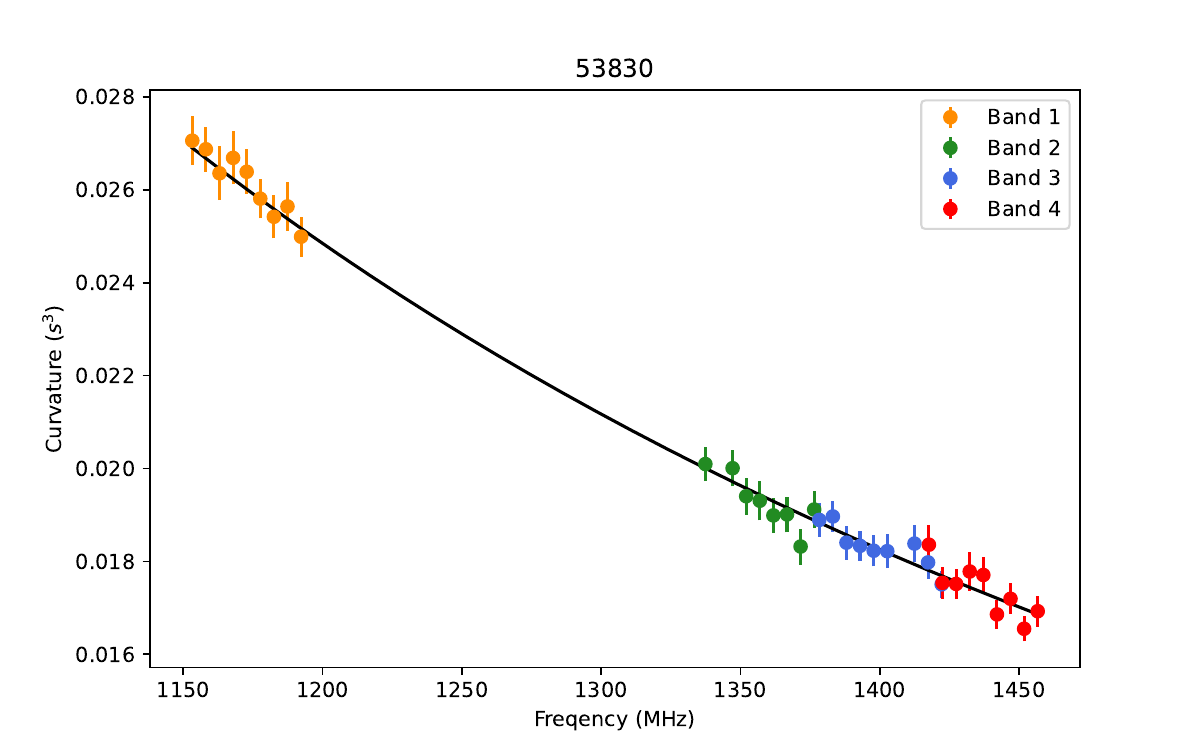}
    
    \end{minipage}
    \begin{minipage}[t]{0.455\textwidth}
        \centering        
        \includegraphics[width=\textwidth]{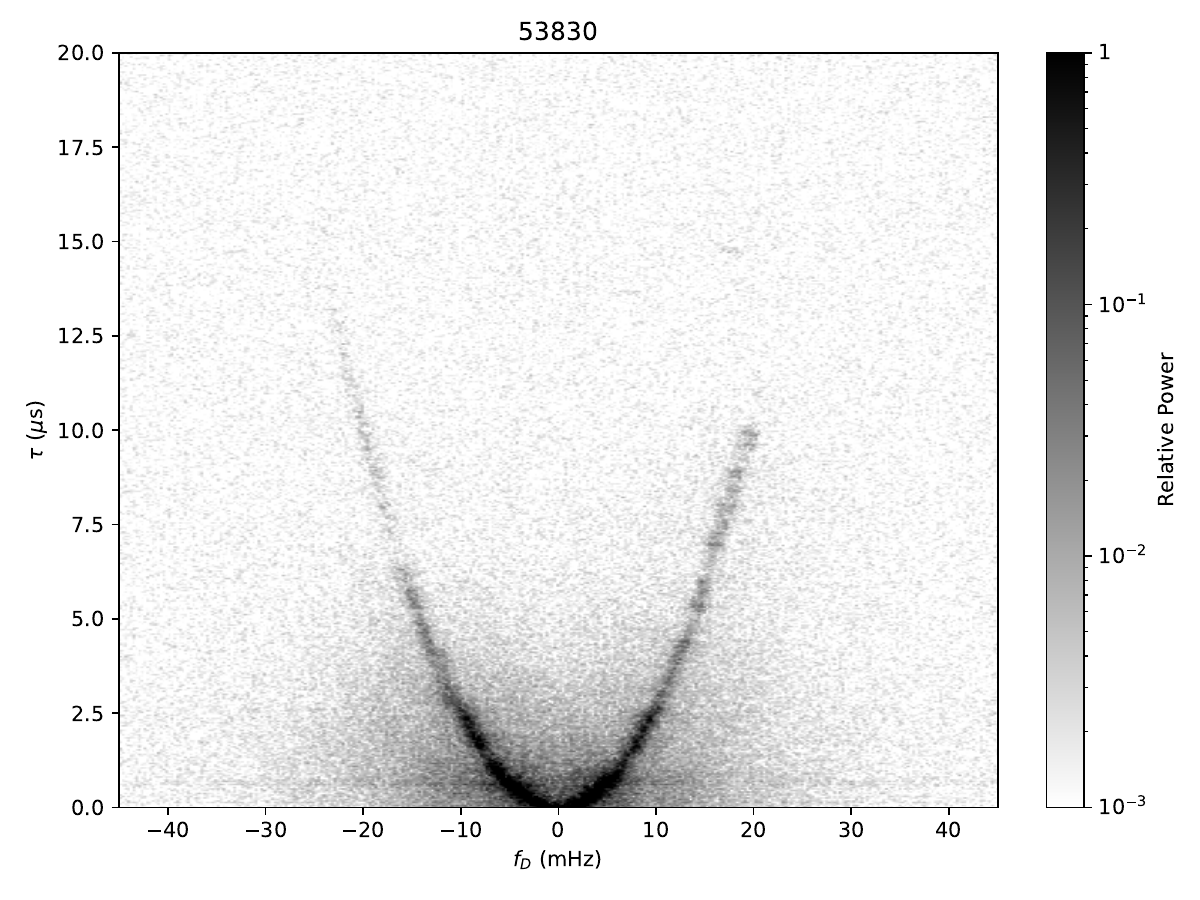}
        \end{minipage}
\caption{The ideal case of curvature measurement and reconstructed wavefield for MJD~53830. The left plot shows the ideal quadratic relationship between curvature and frequency, where each point represents an independent curvature measurement, and different frequency bands are marked by different colors. The global fit over four bands yield a curvature of $(0.0270 \pm 0.0005)\,{\rm s}^3$ at 1175 MHz. The right plot displays the wavefield for MJD~53830 using the best-fitting curvature. In the wavefield, the main arc is clearly modeled by this curvature, allowing for further identification of any additional feature hidden underneath.}
\label{eta v.s. freq}
\label{fig:Perfect_WF}
\end{figure}

\begin{figure}   
    \centering    \includegraphics[width=0.9\textwidth,height=0.58\textwidth]{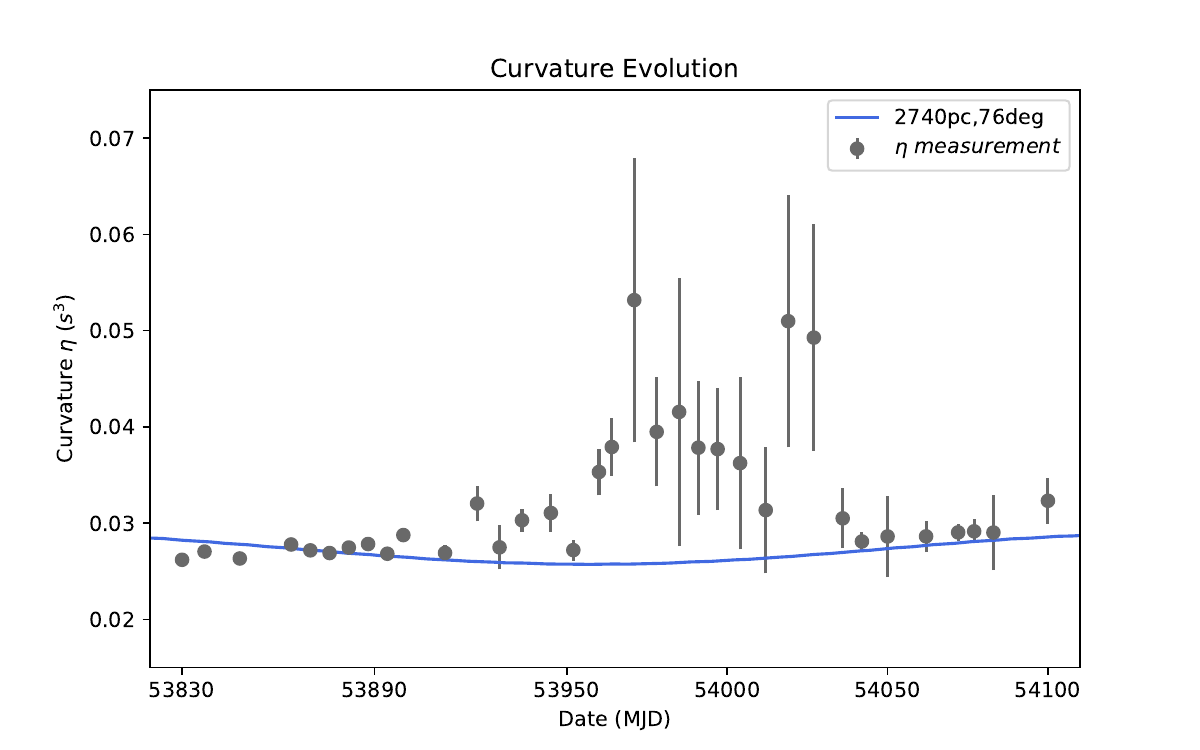}
    \caption{Weekly curvature measurements with the simultaneous fit to all four bands for clean days, before MJD~53888 and after MJD~54050 and the mean and error during the transient event where $\theta-\theta$ become less reliable. Particularly, between MJD~53945 and 54062, the curvatures suddenly jumped to higher values under the interference of secondary lens.} 
    \label{fig:scr1_annual_fitting}
\end{figure}

We repeat this $\theta-\theta$ analysis over all four observing bands to fit for the $\nu^{-2}$ dependence of curvature as can be seen in the left plot in Fig~\ref{eta v.s. freq}. These daily measurements are combined to show the evolution of the curvature over the course of the observing run in Figure~\ref{fig:scr1_annual_fitting}. At the beginning and end of the observing run, the curvature seems to be broadly consistent, but abruptly rises then returns to the baseline value between MJDs 53978 and 54012. This raises an important question: what could cause this sudden and unusual change?

Curvature is a function of the velocities of and distances to the pulsar, ISM, and Earth. Assuming a single screen at a fixed distance, Earth's annual motion is the only factor that can account for the curvature variation. However, the 69-day duration of the curvature change, from MJD~53950 to MJD~54019, seems too short to be caused by the annual variation from the Earth. Moreover, since this epoch matches with the observed passage of the secondary feature through the conjugate spectra, we propose that this change in curvature is caused by the secondary lens.

\subsection{Annual Fitting}\label{sec:Annual Fitting}
Annual fitting is commonly used to determine the distance and orientation of screens, but only when the observation period spans a good fraction of a year or more \citep{Ricket2014,2020ApJ...904..104R}.
Curvature is a function of the screen distance and orientation, and the relative velocity between objects. In our case, the Earth and pulsar velocities are known. Compared to the pulsar motion the screen velocities are usually negligible, so here we assume all the screens have zero transverse velocity. As a result, the curvature value only depends on the distances and orientations of the screens.

\begin{figure}
\centering        
\includegraphics[width=0.8\textwidth]{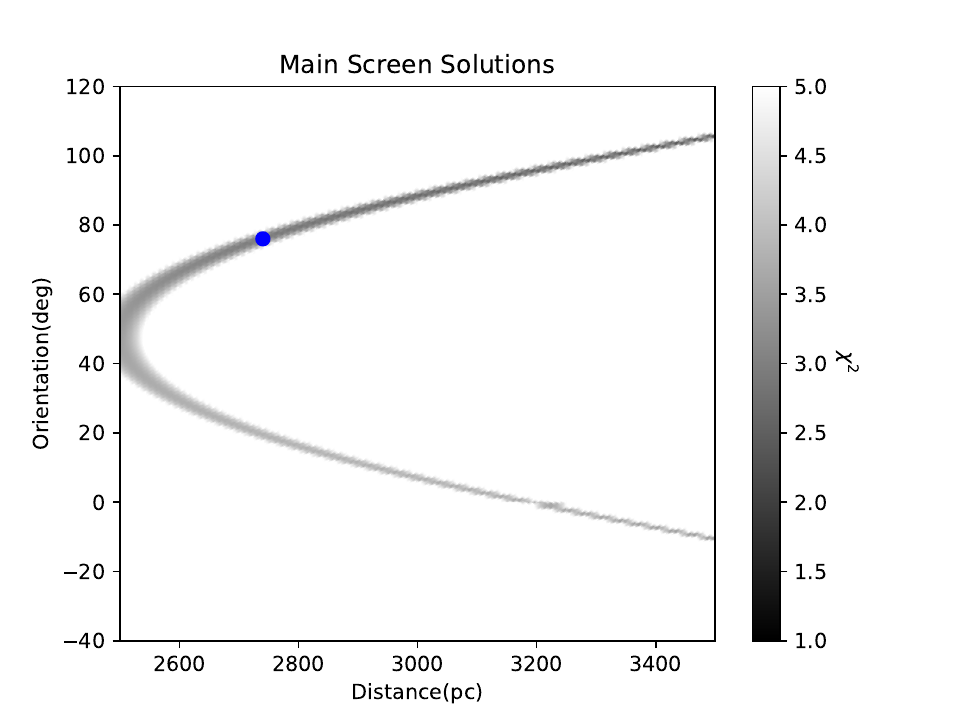}
\caption{Reduced $\chi^2$ of grid search over all possible screen distances and orientations between the pulsar and Earth. Distances below $2500~\rm{pc}$ are strongly excluded because those distances result in imaginary screen images on the sky (see Equation~(\ref{eq:quadratic eqn})).
The blue point is the solution to the main screen we selected because of an acceptable distance and a reduced $\chi^2$ = 2.88.  The annual variation curve of this solution captures the trend of curvature change measured from the data( see Figure~\ref{fig:scr1_annual_fitting}).}
\label{fig:scr1_chi_square_9_5}
\end{figure}

In the B1737+13 data, the total observation period spans nine months, with interference from the secondary lens lasting nearly five months. Therefore, only the first nine epochs and the last five epochs are sufficiently clear for annual fitting of the main screen, leaving a significant gap in the required one-year span. 
Although the data are insufficient for annual fitting, they still provide useful constraints on the main screen solution. 
We ran a grid search over all possible screen distances between the pulsar and Earth (0 to 4176 pc) and orientations (0 to 180\degree) and then calculated the $\chi^2$ of the resulting curvature models, with results shown in Figure~\ref{fig:scr1_chi_square_9_5}.

In general, for a given distance there are two screen orientations, mirrored about the effective velocity, that will produce the same curvature as seen in the two arms of Figure~\ref{fig:scr1_chi_square_9_5}. The degeneracy between these solutions is broken by the slight change in the direction of effective velocity over the course of the year; in our case there is a preference for the upper arm.

We get the main screen solution by doing a least-squares grid search
Figure~\ref{fig:scr1_annual_fitting} by $\chi^2$. The screen distance with the lowest reduced $\chi^2$ occurs at about 3830~pc.
However, that is over 90\% of the distance from the Earth to the pulsar. 
For scattering events, a lens that is too close to either the source or the observer is disfavored because the scattering angle will be too large.
Thus, there is a bias for screens that are located roughly halfway between the pulsar and the Earth.
Hence,  distances less than 3000 pc are  preferable. 
However, from Figure~\ref{fig:scr1_chi_square_9_5}, a distance around 2600 pc has a reduced $\chi^2$ values greater than 3. 
Actually, reduced $\chi^2$ decreases while distance increases. 
Therefore, we chose a distance of 2740 pc and an angle of 76\degree, with reduced $\chi^2 = 2.88$, because its annual variation curve is best fit with the trend of curvature change in Figure~\ref{fig:scr1_annual_fitting} compared to other distance solutions closer to halfway. 

\subsection{Secondary Feature}  \label{sec:secondary-feature}
We focus our analysis of the secondary lens on MJD~53985 to 54012 because those days have the most distinct secondary features. 
In the conjugate spectrum and wavefields, features are expected to enter at negative Doppler shifts, move along their parabolic arcs, and then exit again on the right. Tracking the motion of these features can be helpful in understanding the lenses that formed them. However, when the moving feature gets close enough to the origin, it can be difficult to separate features on different arcs as they overlap as in the left of Figure~\ref{SS53830_54012}. In order to analyze the feature in such cases, we need to separate the signals from different screens. In order to do this, we applied the phase retrieval technique developed by \cite{Baker2022} to recover the underlying wavefield and conjugate wavefield. This collapses the arclets onto the main arc helping to localize features in $\tau$ and $f_D$ space as seen in the right plot of Figure~\ref{fig:Perfect_WF}.

The main arc curvature is needed for phase retrieval to produce a clear wavefield. Based on the main screen annual variation model from \S~\ref{sec:Annual Fitting}, the curvature of the main arc during MJD~53978 to 54012 is 0.026~${\rm s}^3$. This value is then used for phase retrieval to clean up the inverted arclet of the main arc. The resulting wavefields for MJD~58985 and 54012 are shown in Figure~\ref{fig:53985_54012_eyes_fitting} with the secondary feature marked.

Phase retrieval helps us to distinguish between the main arc and extra features. By focusing on the signals inside the small region, we can measure the secondary curvature without interference from the main arc. We had tried the $\theta - \theta$ and Hough transform used in \cite{Reardon:2020sgt} to measure the secondary curvature. The Hough transform finds the best-fitting curvature by adding up all the pixels along the given curvatures. After trying different curvature values, it uses the curvature with the largest power. Unfortunately, in our case, the power of the secondary lens was too weak so these techniques did not work. Moreover, the flux fraction of the moving feature over the whole power of the conjugate spectrum, is approximately 10\%. Obviously, the secondary lens was not bright enough, resulting in failure of both the Hough transform and the $\theta - \theta$ method.

\subsection{Approximate Curvature} \label{sec:secondary-curvature}
\begin{figure}
    \includegraphics[width=\textwidth]{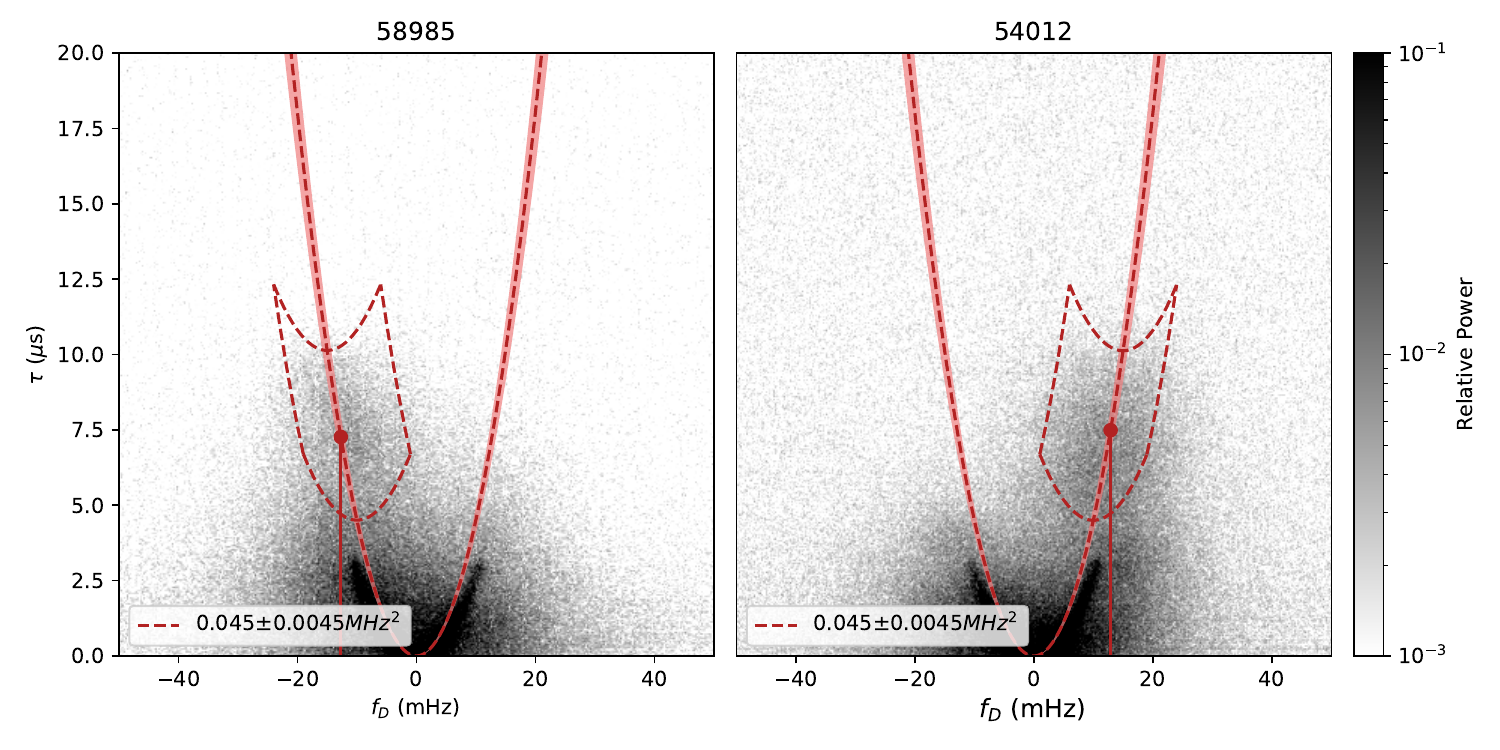}
    \caption{Wavefields for MJD~58985 (left) and MJD~54012 (right), showing the modeled main arc and the identified moving feature. The moving feature, which is the signal from the secondary screen, is marked by the red dashed boxes, with $\eta = 0.045 \pm 0.005$ $\text{MHz}^2$, passing through the center of the feature. The 10\% error comes from eyes is marked by the light red shaded region. At MJD~53985, the moving feature is located around $f_{\rm D} \approx -12 \pm 3$ mHz. By MJD~54012, the feature has shifted to the right side, appearing at approximately $f_{\rm D} \approx 13 \pm 3$ mHz. This displacement ($\Delta f_{\rm D} \approx 25$ mHz over 27 days) aligns with the expected motion of the secondary screen, as predicted by the time derivative in Equation~(\ref{time_derivative}).}
    \centering
    \label{fig:53985_54012_eyes_fitting}
\end{figure}

Since both the Hough transform and $\theta-\theta$ methods failed to recover a curvature for the secondary arc, we instead attempt to fit the arc by eye. For both MJD~58985 and 54012, arc curvature of $0.045~\rm{s}^3$ appears to pass through the center of the feature as seen in Figure~\ref{fig:53985_54012_eyes_fitting}. We therefore use this as the curvature of the secondary arc through the event with an approximate error of $.005~\rm{s}^3$. We can further justify this curvature by noting that, for the one-dimensional screen assumption, the time derivative of $ f_{\rm D}$ for any image on the arc is given by
\begin{equation}\label{time_derivative} 
    \bigdot{f_{\rm D}} =\frac{1}{2 \eta \nu} = 0.82 \pm 0.08 \; {\rm mHz\ day}^{-1}
\end{equation}
as discussed in \cite{Sprenger2022}, \cite{Marthi2021} and \cite{Main2020}.

For an arc with curvature $0.045~\rm{s}^3$, features are expected to move 22 mHz during the 27 days from MJD~58985 and 54012. In Figure~\ref{fig:53985_54012_eyes_fitting}, the distance between the two moving feature is about 25 mHz, which is consistent with the value we got from the time derivative. In order to strengthen the assertion that these features come from the same region on the secondary lens, we note that features at similar $\tau$ have similar offsets from the line of sight, and so should produce comparable magnifications if they have the same lensing potential. The two features used here had $9.7\% \enspace \textrm{and}\enspace 10.0 \%$ of the total flux, respectively. These values were obtained by summing the power within the red region in Figure~\ref{fig:53985_54012_eyes_fitting} and dividing by the entire conjugated spectrum. We are, therefore, confident in our estimate of the secondary screen curvature.

\subsection{The image on the sky}\label{sec:image_ont_the_sky}
Converting the scintillation arc into an image on the sky is critical in the measurement of secondary screen size, which is the key to assessing extreme scattering events, a motivation of this work. However, the distance and orientation of both the primary and secondary screens are needed for the image. In \S~\ref{sec:Annual Fitting} we got the solution for the main screen by fitting the curvature variation. For the secondary screen, since its signal is not clear enough to have a precise curvature measurement and since it is not in the line of sight long enough, the annual-fitting failed to find the distance for it. Instead, we note that for the B0834+06 event (see Appendix~\ref{sec:Appendix_B}), the images formed on the secondary screen remained fixed on a small region of the sky as the pulsar moved. Therefore one potential approach is to map the images from the wavefield onto the sky for different screen parameters and search for a solution that keeps them stationary on the sky over the course of the event. 

The transformation from wavefield space is simply the inverse of Equations~(\ref{eq:tau}) and (\ref{eq:fd}). This corresponds to the intersection of a circle on the sky determined by $\tau$ and a straight line determined by $f_D$. Assuming a known pulsar distance, the free parameters are the screen distance and screen velocity. Since we expect the feature to remain fixed on the sky, we assume a screen velocity of $0~\rm{km}~\rm{s}^{-1}$. This leaves us with screen distance as the only parameter to search over. An intrinsic limitation is that as distance of the screen decreases, the lines of constant $f_D$ tend to spread out; hence, for a given $(f_D,\tau)$ pair there is a minimum distance below which there is no real solution exists. 

We used the points marking the top and bottom of the secondary screen feature in Figure~\ref{fig:53985_54012_eyes_fitting}.
Then, employing
Equations~(\ref{eq:tau}) and (\ref{eq:fd}), we get
\begin{equation}
    a \theta^2 + b \theta +c = 0,
    \label{eq:quart_theta}
\end{equation}
where a, b and c are a function of $V_{\rm eff}$ and $d_{\rm eff}$:
\begin{equation}
   a = 1+\frac{V_x^2}{V_y^2} \quad b = -2\lambda f_d \frac{V_x}{V_y^2}  \quad c = \frac{\lambda^2 f_d^2}{V_y^2}-\frac{2\tau c}{d_{eff}}.
    \label{eq:theta_coefficient}
\end{equation}

\begin{figure}   
    \centering    \includegraphics[width=1\textwidth]{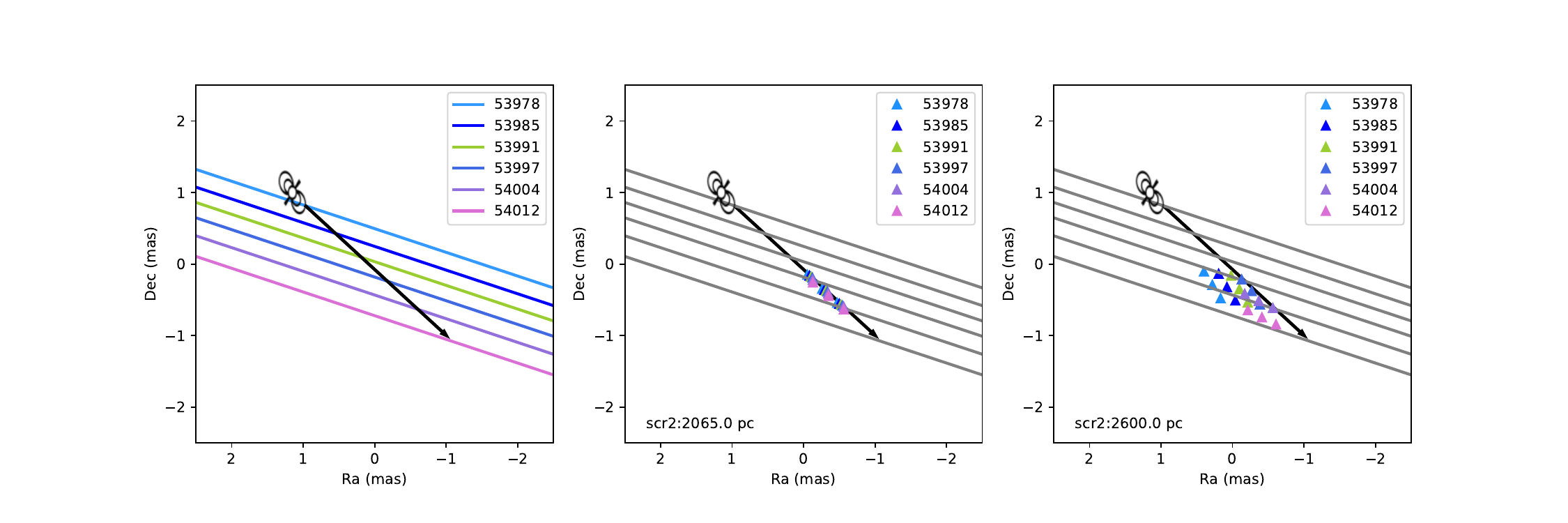} %
    \caption{Images with secondary screens at different distances. Different colors represent images from six dates during the transit of the secondary screen: MJD~53978, 53985, 53991, 53997, 54004 and 54012. Main screens are represented by straight lines, while secondary screens are marked by triangles. The left plot shows the main screen moving with the pulsar. The middle and right plots show the corresponding features at different dates; however, the features are more scattered in the right plot than the middle. In the correct scenario, as the pulsar moves along the black trajectory, the main screen moves with it, while the secondary screen remains fixed at the same position on the sky resulting in secondary features that line up along the pulsar motion. Note: the black vector represents the pulsar parallax over the given period, which differs from $\vv{V}_{\rm eff}$ in previous equations. In B1737+13, the direction of the pulsar parallax is nearly the same as $\vv{V}_{\rm eff}$, as both vectors, parallax and $\vv{V}_{\rm eff}$, are primarily determined by the pulsar’s motion, though with different weighting factors. The displacement test in the lower panel uses data from six dates to measure displacements between data and prediction with different solutions. At a distance of 2065 pc and an orientation of 48\degree, the secondary screen shows the minimum displacement.} 
    \label{fig:six_day_skymap}
\end{figure}

For a real image on the sky, $\sqrt{b^2-4ac} $ needs to be greater than zero, and this forces the distance between the secondary screen and Earth to be more than about 2060~pc. %
Figure~\ref{fig:six_day_skymap} shows the image with real secondary screens at different distances. Setting MJD~53985 as the origin of the image as the pulsar moved, the entire wavefield moved with it since the origin of wavefield is the pulsar itself. 
Therefore, the origin of MJD~53991 is at the position of the pulsar seven days later than MJD~53985. 
For the secondary lens we analyzed for B0834+06, the points remained fixed on the sky as the pulsar moves, since the lens's motion is negligible compared to the pulsar motion on the sky. 
Thus, the features marked by triangles should be overlapping across different dates. 
The left plot in Figure~\ref{fig:six_day_skymap} shows the trajectories of points on the main screen.
We ran all possible distances of the secondary lens, from 2060 pc to 3400 pc, to see whether there was a distance that can make the two features coincide. 
The displacement test presented in the center panel of Figure~\ref{fig:six_day_skymap} shows that the distance that can make this happen has a distance of 2065~pc with an orientation of 48\degree.

It is reasonable that the secondary screen is located at a distance of 2065 pc with a 48\degree orientation. The wavefields in Figure~\ref{fig:WF_and_Simulation_without_yr} show the entire transient event of the secondary screen crossing the line of sight to the pulsar.  As the secondary screen moved across the line of sight, the pulsar motion must have been directed toward the screen in the image, implying that the screen orientation must be aligned with the pulsar moving direction. For B1737+13 during the observation period, the pulsar moved toward 48\degree , which consistent with the angle from the minimum displacement test.

Additionally, we applied this technique to B0834+06 data to verify whether it works on other data sets. B0834+06 is a VLBI data, so its distances and orientation of both screens are clear. This displacement technique successfully came up the same distance for the both screens. Further details of the analysis will be provided in the Appendix~\ref{sec:Appendix_B}.

\subsection{Possible Physical Picture}\label{sec:Possible Physical Picture}
\begin{figure} [htbp]  
    \centering
    \makebox[\textwidth][c]{%
        \includegraphics[width=1\textwidth]{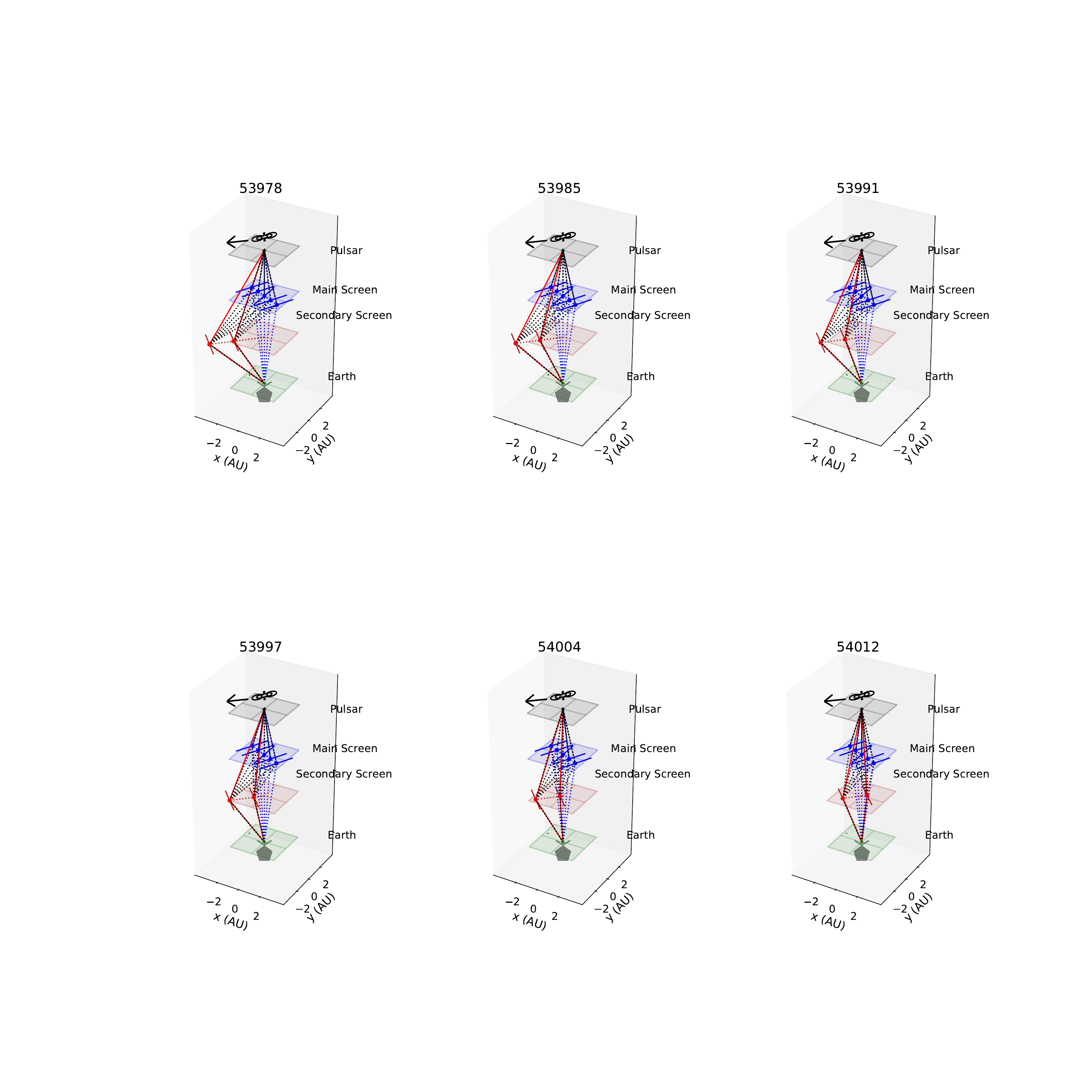}}
    \caption{A schematic diagram of the possible physics picture for B1737+17. The distances between pulsar and screens are not proportional to the reality. In this case, the pulsar and Earth are the only moving object in this system, and the screens remain static. Specifically, the primary screen situated at a distance of 2740 pc is of big size, resulting in extending over the entire observed sky. Since the telescope on Earth always points to the target, the only moving component within this system is the secondary screen located at 2065 pc. This depiction helps conceptualize the dynamics of the system, highlighting the interaction between the pulsar and the moving screens during observations. Also, those six pictures are corresponding to the wavefields in Figure~\ref{fig:WF_and_Simulation_without_yr}.} 
    \centering
    \label{physics_picture}
\end{figure}

Based on the distance and orientation obtained in the previous sections, we proposed a possible alignment in Figure~\ref{physics_picture} to explain these data. 
In this picture, the main screen appears significantly larger, extending across our entire field of view. The relatively small secondary screen 
-- which may be a single lens-like structure -- 
appears to be in motion as it passes through our field of view due to the relative motion between objects. Initially, the light from the pulsar was only scattered by the main screen because the secondary screen was too far from the line of sight.
However, at MJD~53978, another screen entered the scene, blurring the main arc and causing scintillation with different curvatures. As the secondary screen approached the direct line of sight from MJD~53985 to 54004, the scintillation arcs became increasingly blurred as signals from different screens mixed together. Eventually, by MJD~54012, the secondary screen gradually moved out of the view, and the transient event ended. 
The scintillation returned to the arclet-dominated behavior caused by a single screen with a predominantly linear set of lenses.

To further illustrate the scenario presented in  Figure~\ref{physics_picture} and the solution for the secondary screen that we got from the displacement test(Figure~\ref{fig:six_day_skymap} and Appendix~\ref{sec:Appendix_B}), we used the Screens Python package \citep{marten_h_van_kerkwijk_2022_7455536} to simulate a series of wavefields, reproducing the transient event that occurred between MJD~53987 and 54012. 
Using as input the image on the sky like in Figure~\ref{fig:six_day_skymap}, we then determined the wavefields. Table~\ref{simulation_parameters} lists all the parameters used in the simulation.
We used the feature positions on MJD~53985 as the initial positions for the secondary screen. For other dates, we adjusted the feature positions by adding or subtracting $\dot f_{\rm D}$ to approximate its location in the sky, where is different to the previous image mapping, and then used them as input for the simulation. For here, the goal is to simulate a series of wavefields for comparison with the data, rather than to determine exact screen solutions in \S~\ref{sec:image_ont_the_sky}. Thus, we use $\dot{f_{\rm D}}$ to approximate feature positions on dates when the feature overlaps with the main screen; 
this was instead of localizing the features from the wavefield in \S~\ref{sec:image_ont_the_sky}.

\begin{figure}[htbp]    
    \label{fig:WF_and_Simulation_without_yr}
    \centering        
    \makebox[\textwidth][c]{%
        \includegraphics[width=1.2\textwidth]{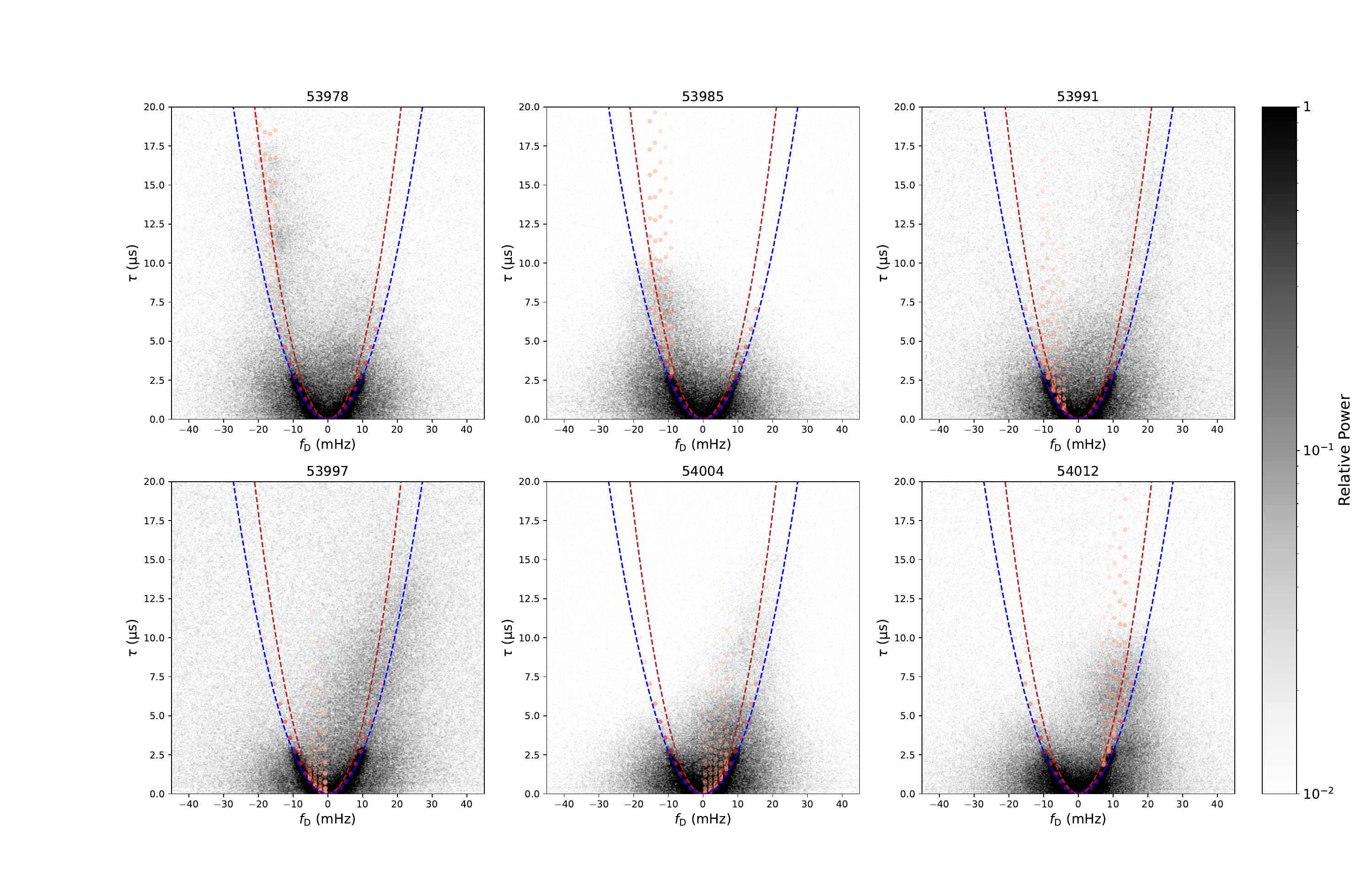}}
    \caption{This is the wavefields with $\eta = 0.026~{\rm s}^3$, which is shown in black and white. The red points are the moving features from the simulation with distances and orientations from \S~\ref{sec:Annual Fitting} and \S~\ref{sec:image_ont_the_sky}, listed in Table ~\ref{simulation_parameters} . Because of the interaction between main and secondary screen, the feature not on the secondary curvature curve(red). The apexes of feature are on the red curve, but the tails are slightly toward the center. Comparing with the simulation and data, they are quite consistent with each other. Therefore, the our parameters are a possible solution for B1737+13. } 
\end{figure}

In the simulation, the primary screen is large enough to cover the entire pulsar  path during the observation, so its position remains fixed over time. For the secondary screen, it is a small static lens, so its relative position to the line of sight changes as the pulsar moves. The corresponding wavefields are in Figure~\ref{fig:WF_and_Simulation_without_yr}. The weak moving feature shows up on the left side and then gradually traverses to the right along the parabolic arc. Eventually, as the secondary screen moves out of  view, the main screen dominates the scintillation again. Also, see the curvature in Figure~\ref{fig:scr1_annual_fitting} for reference.

In this work, we only tracked one moving feature as an example to illustrate the idea of a double-lensing event. However, in reality, the secondary screen in B1737+13 is not just composed of a single moving feature; it likely consists of multiple moving features that pass through the line of sight.
Since this secondary screen is quite weak, we focus on the most obvious moving feature for analysis. Nonetheless, the underlying physics  remains the same.
\begin{deluxetable*}{ccccc}
\tablenum{1}
\tablecaption{The parameters assumed in this possible physics picture.\label{simulation_parameters}}
\tablewidth{0pt}
\tablehead{
\colhead{Object} & \colhead{Distance} &\colhead{Orientation} &\colhead{$\mu_{\delta}$}&\colhead{$\mu_\alpha$}\\
\colhead{}& \colhead{(pc)}& \colhead{(deg)}& \colhead{($\rm{mas}~\rm{yr}^{-1}$)}& \colhead{($\rm{mas}~\rm{yr}^{-1}$)}}
\decimalcolnumbers
\startdata
PSR 1737+13 & 4176 & -- &-22&-20\\
Screen1 & 2740 & 76 &0&0\\
Screen2 & 2065&48&0&0\\
\enddata
\tablecomments{The pulsar parameters are from \cite{Brisken_2003}. }
\end{deluxetable*}

In conclusion, the parameters in Table~\ref{simulation_parameters} are one among several possible solutions for this event in B1737+13.

\subsection{Size of the secondary lens} \label{sec:Size of the secondary screen}
\begin{figure}[htp]
    \begin{minipage}[t]{0.47\textwidth}
        \centering
        \vspace{0pt}
        \includegraphics[width=\textwidth, height=0.4\textheight, trim={0 0 0 0}, clip]{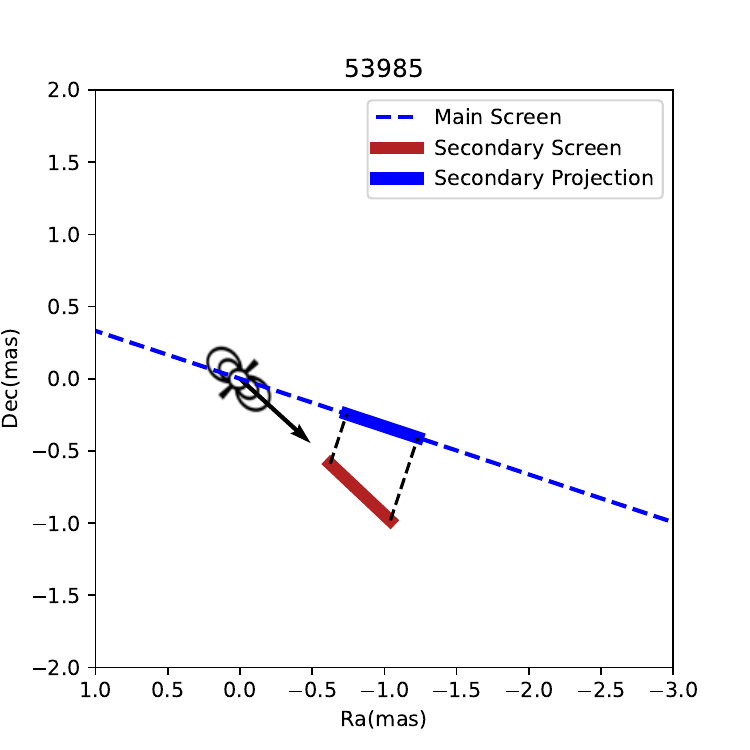} %
        \label{fig:53985_1_0_sky_map_projection}
    \end{minipage}%
    \hfill
    \begin{minipage}[t]{0.53\textwidth}
        \centering
        \vspace{0pt}  
        \includegraphics[width=\textwidth, height=0.40\textheight, trim={0 0 0 0}, clip]{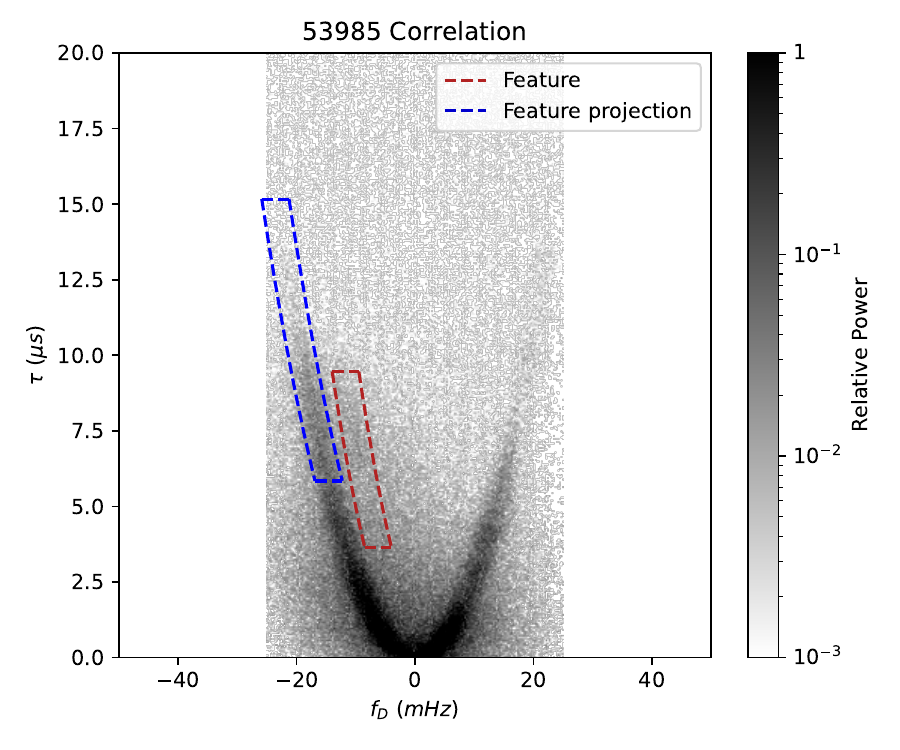}
    \end{minipage}
    \caption{The blue dashed line represent the main screen; the red bold line is the secondary screen with its projection on the main screen marked by blue. The double-lensed image passed both the blue and red bold line region. The corresponding regions on the wavefield are shown on the left. On the right hand panel the blue-dashed box is the secondary screen region on the wavefield; the red-dashed box is the projection of the secondary screen image on the main screen.}    
    \label{fig:flux_correlation}
\end{figure}

The size of the lens is important to the development of a scintillation model. Since we obtained the image in the previous sections, we can use it to estimate the size of secondary screen. Assuming that the flux from the pulsar is isotropic, the flux from the secondary screen passes through both screens.
In contrast, the power in the main arc only passed through the main screen. 
Since the total magnification is the product of the individual screen magnifications, we project the feature of the main screen image and divide those two powers resulting in the magnification of the secondary screen. 
Then, we apply Equation~(\ref{eq:magnification}) to calculate the size of the lens\citep{Simard2018,Zhu2023}. 
From surface brightness conservation, the magnification is
\begin{equation}
    \mu = \frac{d\theta}{d\beta},
    \label{eq:magnification}
\end{equation}
\noindent where $\beta$ is the largest  bending angle to the source, and $\theta$ is the angular position of the image. 
To determine the empirical magnification $\mu$, in the right panel of Figure~\ref{fig:flux_correlation} we measure the flux in the dashed red box and divide that by the flux in the dashed blue box.
Depending on the boundaries we employ, this magnification is in the range $\mu = 0.3$ -- 0.9.
With a measured value of  $\beta = 1.9$~mas, the angular width of the secondary lens is then $\theta' = 0.6$ -- 1.7~mas and its transverse size is in the range  $w' = 1.2$ -- 3.2~au.
%The size of this secondary lens is then  
Because the feature is  fuzzy in the wavefield, it is challenging to define its boundary to get an exact flux of the feature. 
However, this calculation provides a rough size of the lens, confirming that the size is reasonable for an interstellar lens.

%where $\beta$ is the true angular position of a source (relative to the lens center)  and  $\theta$ is the angular position of the image. 
%For a double-lensed case, the image from the secondary screen was lensed twice, so the power of a feature on the wavefield should be divided by its projection on the main arc. Hence, the brightness of the feature should be $\mu = 0.54$, and  
%$\beta = (1.13 \pm 0.21)$~mas, so the angular width $\theta = (0.65 \pm 0.12)$~mas. The size of this secondary lens is $\theta' = (1.34\pm 0.25)$~mas where the error comes from the feature localization. Also we calculated the size on another day, MJD~54012, where the feature is on the other side of the arc with roughly the same $\tau$. The physical size of the secondary lens at this epoch is $\theta' = (0.73\pm 0.15)$~mas. However, because the feature is  fuzzy in the wavefield, it is challenging to define its boundary to get an exact flux of the feature. Therefore, the calculation here  provides a rough size of the lens, confirming that the size is reasonable for an interstellar lens.

Since the signals from the main and secondary screens are highly intertwined, it is challenging to measure the brightness of the feature by simply defining a specific region for it. To address this, we divided the data into two sets based on even and odd time samples and then correlated the signals. 
If a true signal from the secondary screen is present, the amplitude difference between adjacent pixels should be small because the shift is much less than the scintillation time scale, as shown in Figure~\ref{fig:53830_54012_DS}. However, if the data are noise-like the correlation will approach zero. 
The correlation plot is shown in Figure~\ref{fig:flux_correlation}. 
This approach amplifies the signal of the secondary screen on the wavefield and makes it easier to  identify the secondary feature.

\section{Connection to Extreme Scattering Events}
\label{sec:ESE}
Several years before scintillation arcs were recognized as a general phenomenon, \citet{rlg97} reported the occurrence of an ESE in data from the pulsar B0834+06, including a secondary spectrum analysis.
The next reported example of an ESE in pulsar dynamic and secondary spectra 
was the ``millisecond feature'' discovered in the same pulsar during a VLBI  observation \cite{Brisken2010} and further tracked over several weeks in \cite{Zhu2023} using serendipitously adjacent Arecibo-only observations. 
There are several key pieces of evidence that suggest a connection between these structures and ESEs. First, the angular size of the feature must be large enough to cover a distant quasar passing behind it. The millisecond feature was approximately $1.5\pm0.5~\rm{mas}$ compared to the $0.03~\rm{mas}$ of typical images on the main screen. 
The lens investigated here is approximately  0.6 -- 1.6~mas, which is about 1.2 -- 3.2~au at the distance of the secondary screen, much closer to the millisecond feature size.
Another characteristic is the large bending angle of the feature. 
The millisecond feature was visible at $24~\rm{mas}$ at approximately $320~\rm{MHz}$ with no other images forming nearby on the sky.  Correcting for the $\nu^{-2}$ scaling of the bending angle with frequency, the secondary screen reported here could scatter the signal by at least as much. Unfortunately, the feature first appears at the very top of our conjugate spectrum, and observations with higher frequency resolution may have been able to detect it from even further out. We conclude that the secondary screen in this work is consistent with being the same type of structure as in \cite{Zhu2023}, but the close proximity of the two screens in conjugate spectrum space makes them more difficult to isolate and study, and so we cannot rule out the possibility that a more typical scintillation screen is responsible for the second arc.

\section{Conclusion}
\label{sec:Conclusion}
In this work, we present a new analysis of archival observations of PSR B1737+13 for which a transient feature is observed passing through the conjugate spectrum over the course of several months. During this period the arc transitions from a clean arc with inverted arclets to one significantly more blurry, making curvature measurements for the feature difficult. Phase retrieval allows us to isolate a main arc as well as several isolated images formed by a secondary lensing structure.

With limited data, annual fitting provides a constrained solution pool for the primary screen. We selected a solution near the midpoint, as the diffracted angle becomes larger if the screen is too close to the pulsar. The secondary feature appears similar to the millisecond feature seen in \cite{Brisken2010}, which is believed to be related to the
ESEs
seen in quasars. 
We present a new technique for measuring the distances to such structures, 
minimizing the variance of the average on sky position of images over multiple days as a function of screen distance.
As described in Appendix~\ref{sec:Appendix_B}, we validate this technique on the millisecond feature before applying it to our new observations to get a distance and an approximate on-sky position of the structure. The solutions for the two screens were then used to simulate a two-lens system where the pulsar signal can be deflected by one or both screens on its way to Earth.

These simulations agree well with the observed conjugate spectra, including the blurring of the main arc and the positions of the doubly lensed images in $\tau-f_D$ space. To help understand the structures in these double lensing models, we introduce the concept of an interaction arc and interaction curvature to describe the pattern of the double-lensed image in the conjugate spectrum, which moves the feature slightly off from the arc expected of an image lensed by only a single screen.
The interaction arc concept will be important for future studies of these doubly lensed systems. 
For systems without appreciable double lensing interaction between the two screens, the number of images in the conjugate wavefield will be the sum of the images formed by each independently. However, points from double lensing scale with the product of the number of images on each screen. This can greatly complicate the wavefield,especially as two or more of the resulting arcs may intersect. More details on the interaction curvature are given in Appendix~\ref{sec:Appendix_A}.

With these identified solutions, we were also able to estimate the size of the secondary lens, finding  it to be au-scale in size, which is a typical ISM lens size. 
% However, it is an ordinary scintillation screen, not the screen causing extreme scattering events(ESEs). 
In these data we do not see any extremely bright and compact features such as the screen edges that can cause ESEs like those described in \cite{Zhu2023}. In fact, the secondary screen occupies the line of sight for more than 12 weeks. In this study we simply pick one clear feature on the secondary screen for analysis, and its transit time is about 6 weeks. Other features on the secondary screen were either too weak or  interacted strongly with (were confused by) the main screen, making them difficult to identify and track.

Although the lens causing this transient may not be an ESE candidate -- we do not have conclusive evidence for or against that hypothesis -- 
we have studied a double-lensing case in detail and propose a new method to identify the screen distance and orientation. With anticipated VLBI data on B1737+13, we may be able to further resolve the screen geometries and internal structure.

\section*{Acknowledgments}
%\begin{acknowledgments}
%\begin{center}
%{\large Acknowledgments}
%\end{center}
\noindent Support for this research at Oberlin College was provided by an NSF Physics Frontiers Center award (2020265) to NANOGrav and through an NSF RUI grant (2009759, PI Stinebring).
The authors would like to acknowledge the funding support of the Canadian Institute for Advanced Research Fund, the Natural Sciences and Engineering Research Council of Canada (funding reference number RGPIN-2019-06770, ALLRP 586559-23), the Ontario Research Fund -- Research Excellence (ORF-RE Fund), and AMD AI Quantum Astro (Fund number 473735).
DS also thanks several Oberlin College students who worked on earlier versions of this research; in particular: Jakob Faber, Stella Ocker, and Hengrui Zhu.
%\end{acknowledgments}

\facilities{Arecibo}
\\
\\
{\em\large Objects:} 
\object{PSR B1737+13}

\FloatBarrier

\appendix 
\section{Interaction Arcs}
\label{sec:Appendix_A}
\subsection{Definition}
\begin{figure}
    \centering
    \includegraphics[width=\columnwidth]{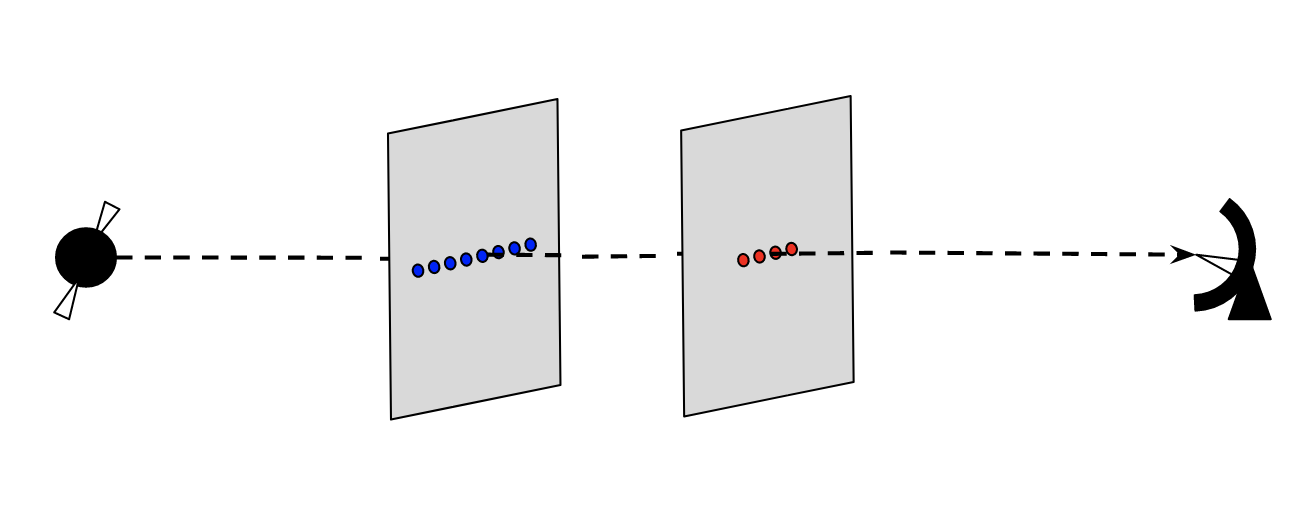}
    \caption{A schematic of a pulsar-screen system with a primary scattering screen (blue images) and a secondary scattering screen (red images), with the primary scattering screen located closer to the pulsar. The primary screen has a larger extent (more images) than the secondary screen).}
    \label{fig:interactcurvesitu}
\end{figure}
\begin{figure}
    \centering
    \includegraphics[width=0.9\columnwidth]{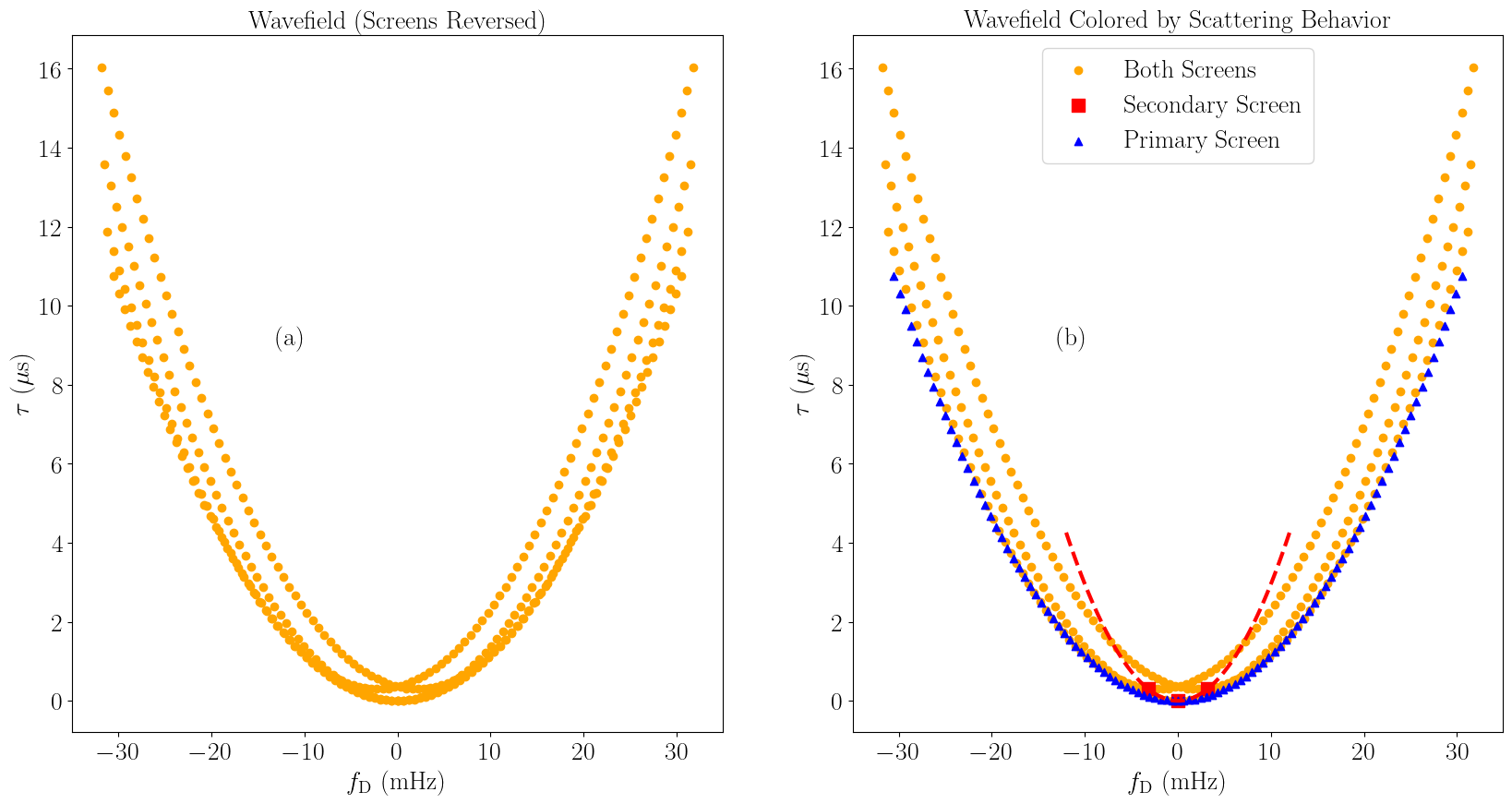}
    \caption{The simulated wavefield from a situation like that in Figure \ref{fig:interactcurvesitu}, created using the Screens package \citep{marten_h_van_kerkwijk_2022_7455536}. The primary screen is composed of 101 points, and the secondary screen is composed of five points. In the left panel, the composite wavefield from scattering due to both screens and scattering due to the primary screen only is shown. In the right panel,
    the wavefield from the interaction of the two screens is shown in orange, the wavefield from scattering due to the primary screen only is shown in blue,
    and the wavefield from scattering due to the secondary screen only is shown in red and connected by a dotted parabola to indicate the curvature of the secondary screen.}
    \label{fig:interactioncurvaturedefinition}
\end{figure}
\begin{figure}
    \centering
    \includegraphics[width=\columnwidth]{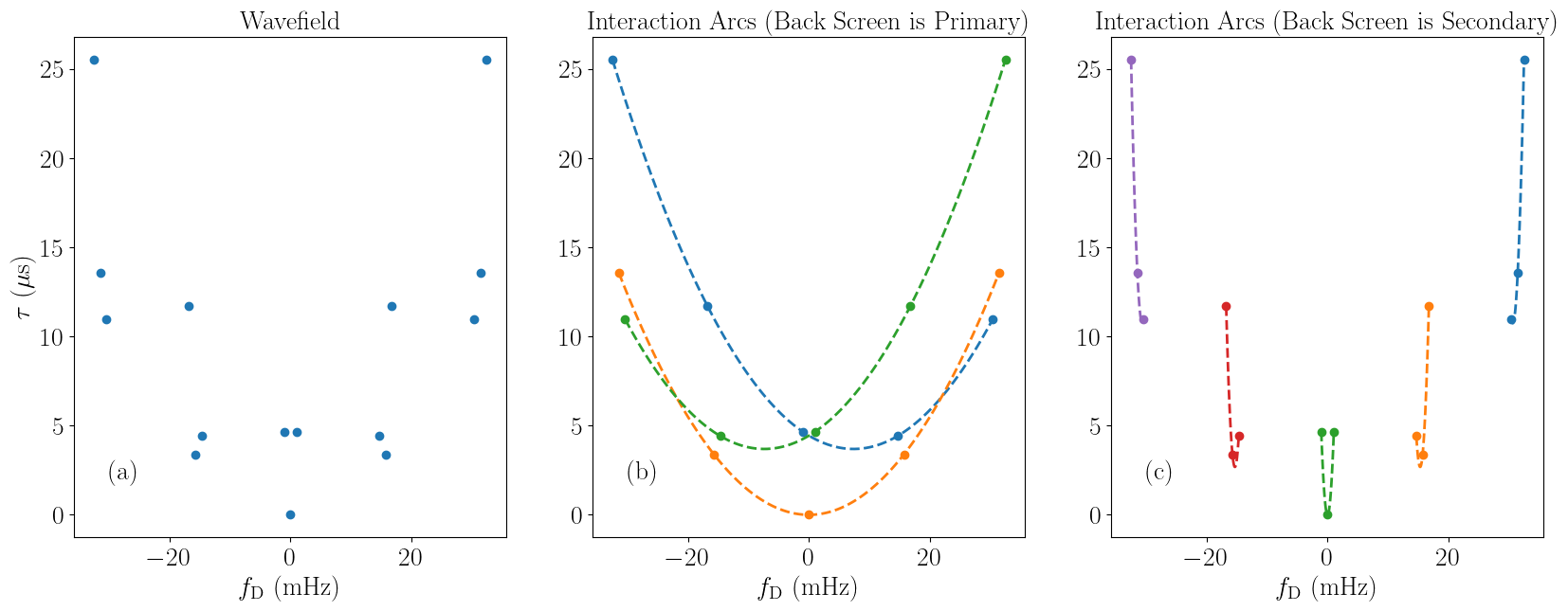}
    \caption{Panel (a) shows the wavefield from the interaction between a screen closer to the observer (the front screen) with three points and a screen farther from the observer (the back screen) with five points. Panel (b) groups wavefield points into three parabolas of five points each, with each point on a parabola being scattered by a common point on the secondary screen. Panel (c) groups wavefield points into five parabolas of three points each, with each point on a parabola being scattered by a common point on the primary screen. Both ways of drawing interaction arcs are essentially equivalent.}
    \label{fig:interactionarcstwoways}
\end{figure}

Here, we introduce the term \textit{interaction arcs} as a way to discuss the wavefield resulting from light scattered by multiple screens en route to the observer.
Consider a situation like that shown in Figure~\ref{fig:interactcurvesitu}, where a primary scattering screen is located closer to the pulsar than the secondary scattering screen, and both screens are composed of discrete scattering points for the sake simulating such a situation using the Screens Python package \citep{marten_h_van_kerkwijk_2022_7455536}. The primary screen is considered ``primary'' in the sense that it is larger in extent and thus contains more scattering points. When we simulate this situation, we arrive at a wavefield shown in Figure~\ref{fig:interactioncurvaturedefinition}. This wavefield shows two scintillation arcs representing points scattered by only the primary screen or only the secondary screen. However, we also see a set of larger scintillation arcs spread across the wavefield that situ on top of the primary arc with their apexes lying on the secondary screen's scintillation arc. These higher curvature arcs represent points in the wavefield from light that has scattered off of both the secondary and the primary screens, and it is this set of arcs that we call \textit{interaction arcs.} For discrete scattering points, if there are $N_1$ points in the primary screen and $N_2$ points in the secondary screen, then there are $N_1N_2$ points in the wavefield from the scattering due to both screens.
\\\indent The definition as described above hides the ambiguity involved in assigning the $N_1N_2$  points in the wavefield to the different parabolas we observe. This ambiguity is demonstrated in Figure~\ref{fig:interactionarcstwoways} where $N_1=5$ and $N_2 = 3$. Parabolas can be drawn linking \textit{either} points scattered by a common point in the primary screen \textit{or} by points scattered by a common point in the secondary screen. The choice of method used to assign points to parabolas is essentially arbitrary for a situation like that in Figure~\ref{fig:interactionarcstwoways}. 
In Figure~\ref{fig:interactioncurvaturedefinition}, however, there are many more points in the primary screen compared to the secondary screen, so the assignment of points to interaction arcs appears more obvious to the eye. Here, we have mainly considered two-screen scenarios where the primary screen is much larger in extend than the secondary screen, leading to the following working definition: \textit{\textbf{Interaction arcs} are parabolic sets of points in the wavefield resulting from images scattered twice: once by the primary screen and once by a common point on the secondary screen. The \textbf{interaction curvature}, $\eta_{\rm interact}$, is the curvature of an interaction arc.}
\subsection{Implications of Interaction Arcs for B1737+13}
\begin{figure}
    \centering
    \includegraphics[width=\textwidth]{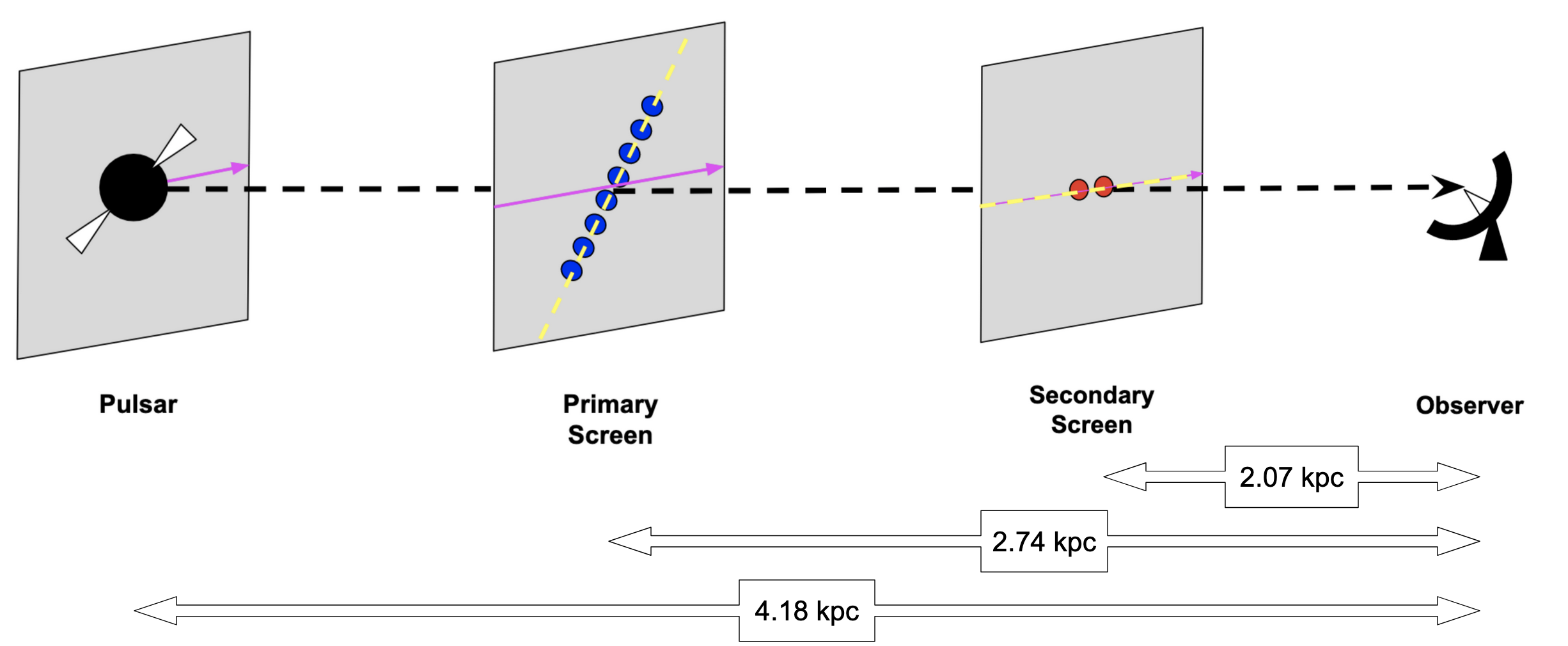}
    \caption{A diagram showing a possible configuration of the B1737+13 system with two scattering screens. The number of points on each screen indicates the relative extent of each scattering screen. The magenta arrows indicate the direction of the pulsar's motion projected onto each screen. The yellow dashed lines indicate the orientation of the line of images on each screen.}
    \label{fig:physics_picture_cartoon}
\end{figure}

The concept of interaction arcs may offer insight into a possible physical picture behind the scattering behavior observed in pulsar B1737+13. While a two-screen model appears to present a likely physical explanation behind the measured curvatures and observed moving features in the conjugate spectrum, two screens acting independently of each other do not explain the observations by themselves. If no light was scattered by \textit{both} the primary and secondary screen, we would expect to see two distinct scintillation arcs with different curvatures in the conjugate spectrum. This type of behavior has been observed in other pulsars; for example, up to six arcs have been identified in pulsar B1133+16 \citep{mckee_2022ApJ...927...99M}. Apart from the beginning and end of observations—where a single curvature dominates—B1737+13 has a conjugate spectrum without well-resolved scintillation arcs. The conjugate spectrum is instead ``fuzzy,'' with the primary arc washed out by diffuse regions of power.
\\
\indent Interaction arcs resulting from two-screen scattering are capable of causing fuzziness in a secondary spectra of. In the wavefield shown in Figure~\ref{fig:interactioncurvaturedefinition} representing the situation in \ref{fig:interactcurvesitu}, for example, the interaction arcs sit superimposed on the top of the primary and secondary arcs, resulting in power distributed in a much larger region in the wavefield than with independent screens. It is possible that a similar situation could also be responsible for the fuzziness seen with B1737+13. The proposed physical picture shown in Figure~\ref{physics_picture} and illustrated in Figure~\ref{fig:physics_picture_cartoon} is quite similar to the situation in Figure~\ref{fig:interactcurvesitu}. The distances and orientations of the primary and secondary screens have been adjusted such that they produce scintillation arcs with curvatures of $\eta_1 = 0.026 \mathrm{\;s^3}$ and $\eta_2 = 0.045 \mathrm{\;s^3}$, respectively.

\section{B0834+06 Data} \label{sec:Appendix_B}
Doing the displacement test on screen images, we successfully constrained the solution for the secondary screen in B1737+13. This provide a new method for single-dished observations that can determine screen properties. However, whether the displacement test can be applied to other cases or if just lucky success in B1737+13, need more examples. To test it, we applied the method to another double-lensed pulsar, B0834+06, for which the screen properties are known from VLBI observation.

\begin{figure}   
    \centering    \includegraphics[width=0.5\textwidth,height=0.4\textwidth]{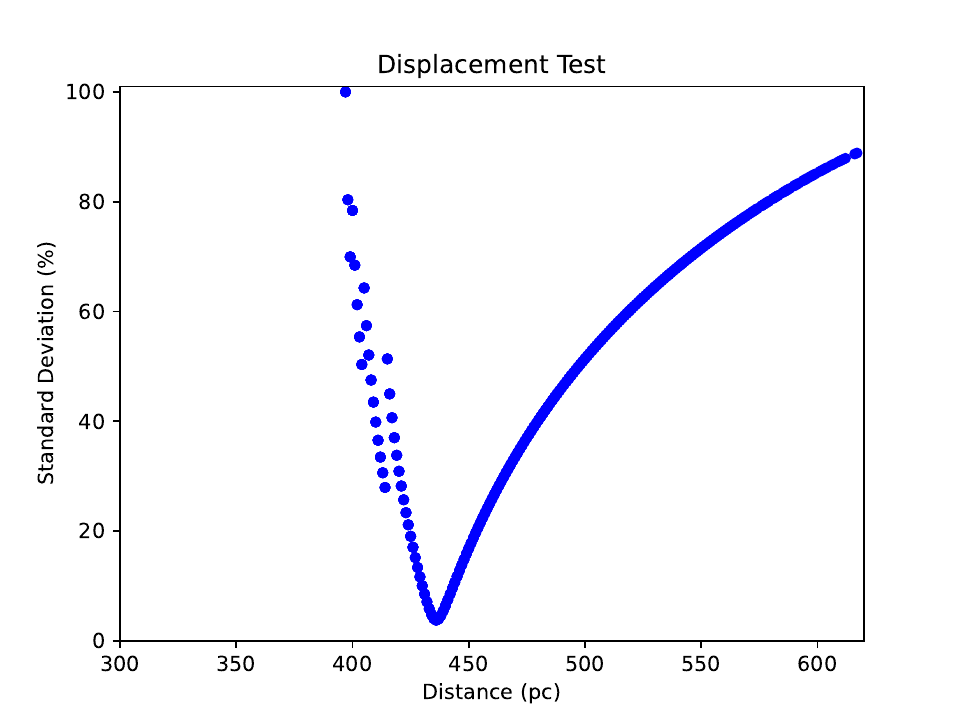}
    \caption{From the displacement test, 435 pc is the minimum displacement between two feature from two different dates, indicating that it is the solution for this system under the frame of static screen on the sky. The distance predicted by this displacement test (435 pc) is close to the distance obtained from VLBI (415 pc). Hence, this example demonstrates that the displacement test is a useful for determining distances in single-dish observations. } 
    \label{fig:Brisken_data_best_solution}
\end{figure}

As mentioned in \S~\ref{sec:image_ont_the_sky}, we tested various distances and orientations that produced the same curvature as observed, and then transform the wavefields into the images. The correct distance should make the features at the same position on the sky as pulsar moves. Therefore, the features with less displacement between two different date is the proper solutions for the system. We take the average of all the points in a feature as its center since there are six points in one, and then measure the standard deviation of the two feature center from different dates. A smaller standard deviation indicates that the features are closer to each other. Hence, there is a special distance can make the features from two different date nearly overlap, as shown in Figure~\ref{fig:Brisken_data_best_solution}. For B0834+06, is at a distance of 435 pc from Earth, as the drop in Figure~\ref{fig:Brisken_data_best_solution}. This result is at  1.8 $\sigma$  comparing to the distance of $415\pm 11$ pc provided by \cite{lpm+16}.

\begin{figure}[htbp]   
    \centering    
    \makebox[\textwidth][c]{%
        \includegraphics[width=1\textwidth]{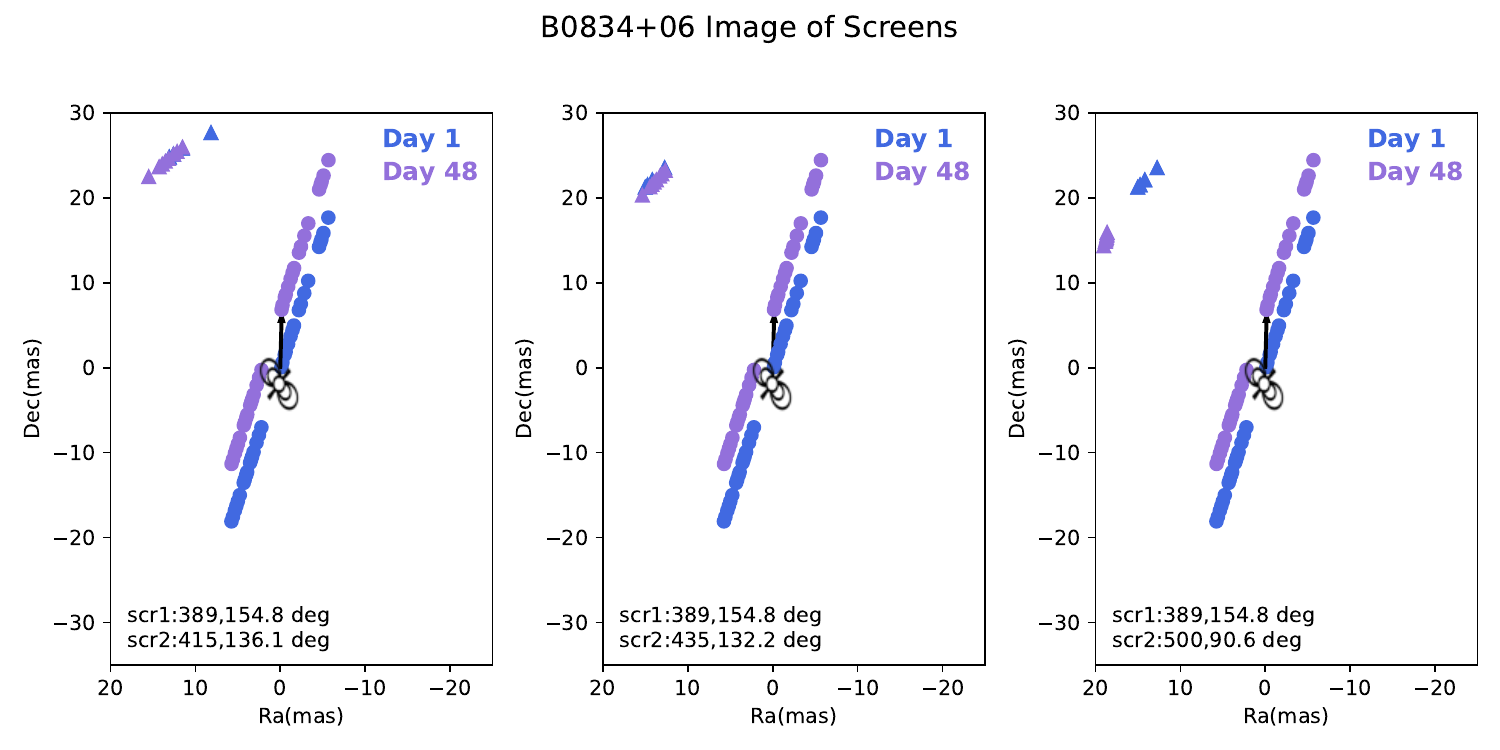}
    }
    \caption{Screen images of B0834+06 with secondary screens at various distances. Blue marks represent day 1, and green marks represent day 48. The main screen at 2740 pc is indicated by round points, while the secondary screen is shown as triangles. $\delta f$ represents the distance between the two features, with normalized displacement shown below. The black vector indicates the pulsar's trajectory over the given period. In the left plot, 415 pc is the distance with the minimum displacement in the displacement test. The middle plot shows the VLBI image with the secondary screen at 435 pc. In the right plot, although this distance and orientation yield the same curvature, the features are far away from each other, suggesting that this distance is unlikely for B0834+06.} 
    \label{scr1_annual_fitting}
\end{figure}

\FloatBarrier
%\bibliography{Refs/yenhua,Refs/scintillation,Refs/modrefs,Refs/psrrefs}{} % use the .bbl file going forward

\begin{thebibliography}{}
\expandafter\ifx\csname natexlab\endcsname\relax\def\natexlab#1{#1}\fi
\providecommand{\url}[1]{\href{#1}{#1}}
\providecommand{\dodoi}[1]{doi:~\href{http://doi.org/#1}{\nolinkurl{#1}}}
\providecommand{\doeprint}[1]{\href{http://ascl.net/#1}{\nolinkurl{http://ascl.net/#1}}}
\providecommand{\doarXiv}[1]{\href{https://arxiv.org/abs/#1}{\nolinkurl{https://arxiv.org/abs/#1}}}

\bibitem[{{Baker} {et~al.}(2022){Baker}, {Brisken}, {van Kerkwijk}, {Main},
  {Pen}, {Sprenger}, \& {Wucknitz}}]{Baker2022}
{Baker}, D., {Brisken}, W., {van Kerkwijk}, M.~H., {et~al.} 2022, \mnras, 510,
  4573, \dodoi{10.1093/mnras/stab3599}

\bibitem[{Baker {et~al.}(2023)Baker, Brisken, van Kerkwijk, van Lieshout, \&
  Pen}]{baker-2023}
Baker, D., Brisken, W., van Kerkwijk, M.~H., van Lieshout, R., \& Pen, U.-L.
  2023, Monthly Notices of the Royal Astronomical Society, 525, 211,
  \dodoi{10.1093/mnras/stad2318}

\bibitem[{{Bogdanov} {et~al.}(2002){Bogdanov}, {Pruszy{\'n}ska}, {Lewandowski},
  \& {Wolszczan}}]{Bogdanov2002}
{Bogdanov}, S., {Pruszy{\'n}ska}, M., {Lewandowski}, W., \& {Wolszczan}, A.
  2002, \apj, 581, 495, \dodoi{10.1086/344169}

\bibitem[{Brisken {et~al.}(2003)Brisken, Fruchter, Goss, Herrnstein, \&
  Thorsett}]{Brisken_2003}
Brisken, W.~F., Fruchter, A.~S., Goss, W.~M., Herrnstein, R.~M., \& Thorsett,
  S.~E. 2003, The Astronomical Journal, 126, 3090, \dodoi{10.1086/379559}

\bibitem[{{Brisken} {et~al.}(2010){Brisken}, {Macquart}, {Gao}, {Rickett},
  {Coles}, {Deller}, {Tingay}, \& {West}}]{Brisken2010}
{Brisken}, W.~F., {Macquart}, J.~P., {Gao}, J.~J., {et~al.} 2010, \apj, 708,
  232, \dodoi{10.1088/0004-637X/708/1/232}

\bibitem[{{Cordes} \& {Rickett}(1998)}]{Cordes1998}
{Cordes}, J.~M., \& {Rickett}, B.~J. 1998, \apj, 507, 846,
  \dodoi{10.1086/306358}

\bibitem[{{Cordes} {et~al.}(2006){Cordes}, {Rickett}, {Stinebring}, \&
  {Coles}}]{crsc06}
{Cordes}, J.~M., {Rickett}, B.~J., {Stinebring}, D.~R., \& {Coles}, W.~A. 2006,
  \apj, 637, 346, \dodoi{10.1086/498332}

\bibitem[{{Fiedler} {et~al.}(1987){Fiedler}, {Dennison}, {Johnston}, \&
  {Hewish}}]{Fiedler1987}
{Fiedler}, R.~L., {Dennison}, B., {Johnston}, K.~J., \& {Hewish}, A. 1987,
  \nat, 326, 675, \dodoi{10.1038/326675a0}

\bibitem[{{Hemberger} \& {Stinebring}(2008)}]{2008ApJ...674L..37H}
{Hemberger}, D.~A., \& {Stinebring}, D.~R. 2008, \apjl, 674, L37,
  \dodoi{10.1086/528985}

\bibitem[{{Jow} {et~al.}(2023){Jow}, {Pen}, \& {Baker}}]{Jow2023}
{Jow}, D.~L., {Pen}, U.-L., \& {Baker}, D. 2023, arXiv e-prints,
  arXiv:2301.08344, \dodoi{10.48550/arXiv.2301.08344}

\bibitem[{{Liu} {et~al.}(2016){Liu}, {Pen}, {Macquart}, {Brisken}, \&
  {Deller}}]{lpm+16}
{Liu}, S., {Pen}, U.-L., {Macquart}, J.-P., {Brisken}, W., \& {Deller}, A.
  2016, \mnras, 458, 1289, \dodoi{10.1093/mnras/stw314}

\bibitem[{{Lyne}(1984)}]{Lyne1984}
{Lyne}, A.~G. 1984, \nat, 310, 300, \dodoi{10.1038/310300a0}

\bibitem[{{Main} {et~al.}(2020){Main}, {Sanidas}, {Antoniadis}, {Bassa},
  {Chen}, {Cognard}, {Gaikwad}, {Hu}, {Janssen}, {Karuppusamy}, {Kramer},
  {Lee}, {Liu}, {Mall}, {McKee}, {Mickaliger}, {Perrodin}, {Stappers},
  {Tiburzi}, {Wucknitz}, {Wang}, \& {Zhu}}]{Main2020}
{Main}, R.~A., {Sanidas}, S.~A., {Antoniadis}, J., {et~al.} 2020, \mnras, 499,
  1468, \dodoi{10.1093/mnras/staa2955}

\bibitem[{{Marthi} {et~al.}(2021){Marthi}, {Simard}, {Main}, {Pen}, {van
  Kerkwijk}, {Vanderlinde}, {Gupta}, {Roberts}, \& {Quine}}]{Marthi2021}
{Marthi}, V.~R., {Simard}, D., {Main}, R.~A., {et~al.} 2021, \mnras, 506, 5160,
  \dodoi{10.1093/mnras/stab1970}

\bibitem[{{McKee} {et~al.}(2022){McKee}, {Zhu}, {Stinebring}, \&
  {Cordes}}]{mckee_2022ApJ...927...99M}
{McKee}, J.~W., {Zhu}, H., {Stinebring}, D.~R., \& {Cordes}, J.~M. 2022, \apj,
  927, 99, \dodoi{10.3847/1538-4357/ac460b}

\bibitem[{Ocker {et~al.}(2023)Ocker, Cordes, Chatterjee, Stinebring, Dolch,
  Pelgrims, McKee, Giannakopoulos, \& Reardon}]{Ocker2023}
Ocker, S.~K., Cordes, J.~M., Chatterjee, S., {et~al.} 2023.
\newblock \doarXiv{2309.13809}

\bibitem[{{Pen} \& {Levin}(2014)}]{Pen2014}
{Pen}, U.-L., \& {Levin}, Y. 2014, \mnras, 442, 3338,
  \dodoi{10.1093/mnras/stu1020}

\bibitem[{{Reardon} {et~al.}(2020{\natexlab{a}}){Reardon}, {Coles}, {Bailes},
  {Bhat}, {Dai}, {Hobbs}, {Kerr}, {Manchester}, {Os{\l}owski}, {Parthasarathy},
  {Russell}, {Shannon}, {Spiewak}, {Toomey}, {Tuntsov}, {van Straten},
  {Walker}, {Wang}, {Zhang}, \& {Zhu}}]{rcb+20}
{Reardon}, D.~J., {Coles}, W.~A., {Bailes}, M., {et~al.} 2020{\natexlab{a}},
  \apj, 904, 104, \dodoi{10.3847/1538-4357/abbd40}

\bibitem[{{Reardon} {et~al.}(2020{\natexlab{b}}){Reardon}, {Coles}, {Bailes},
  {Bhat}, {Dai}, {Hobbs}, {Kerr}, {Manchester}, {Os{\l}owski}, {Parthasarathy},
  {Russell}, {Shannon}, {Spiewak}, {Toomey}, {Tuntsov}, {van Straten},
  {Walker}, {Wang}, {Zhang}, \& {Zhu}}]{2020ApJ...904..104R}
---. 2020{\natexlab{b}}, \apj, 904, 104, \dodoi{10.3847/1538-4357/abbd40}

\bibitem[{Reardon {et~al.}(2020)}]{Reardon:2020sgt}
Reardon, D.~J., {et~al.} 2020, Astrophys. J., 904, 104,
  \dodoi{10.3847/1538-4357/abbd40}

\bibitem[{{Rickett}(1990)}]{ric90}
{Rickett}, B.~J. 1990, \araa, 28, 561,
  \dodoi{10.1146/annurev.aa.28.090190.003021}

\bibitem[{{Rickett} {et~al.}(1997){Rickett}, {Lyne}, \& {Gupta}}]{rlg97}
{Rickett}, B.~J., {Lyne}, A.~G., \& {Gupta}, Y. 1997, \mnras, 287, 739,
  \dodoi{10.1093/mnras/287.4.739}

\bibitem[{{Rickett} {et~al.}(2014{\natexlab{a}}){Rickett}, {Coles}, {Nava},
  {McLaughlin}, {Ransom}, {Camilo}, {Ferdman}, {Freire}, {Kramer}, {Lyne}, \&
  {Stairs}}]{Rickett2014}
{Rickett}, B.~J., {Coles}, W.~A., {Nava}, C.~F., {et~al.} 2014{\natexlab{a}},
  \apj, 787, 161, \dodoi{10.1088/0004-637X/787/2/161}

\bibitem[{{Rickett} {et~al.}(2014{\natexlab{b}}){Rickett}, {Coles}, {Nava},
  {McLaughlin}, {Ransom}, {Camilo}, {Ferdman}, {Freire}, {Kramer}, {Lyne}, \&
  {Stairs}}]{Ricket2014}
---. 2014{\natexlab{b}}, \apj, 787, 161, \dodoi{10.1088/0004-637X/787/2/161}

\bibitem[{{Simard} \& {Pen}(2018)}]{Simard2018}
{Simard}, D., \& {Pen}, U.-L. 2018, \mnras, 478, 983,
  \dodoi{10.1093/mnras/sty1140}

\bibitem[{{Sprenger} {et~al.}(2022){Sprenger}, {Main}, {Wucknitz}, {Mall}, \&
  {Wu}}]{Sprenger2022}
{Sprenger}, T., {Main}, R., {Wucknitz}, O., {Mall}, G., \& {Wu}, J. 2022,
  \mnras, 515, 6198, \dodoi{10.1093/mnras/stac2160}

\bibitem[{Sprenger {et~al.}(2020)Sprenger, Wucknitz, Main, Baker, \&
  Brisken}]{TimSprenger2021}
Sprenger, T., Wucknitz, O., Main, R., Baker, D., \& Brisken, W. 2020, Monthly
  Notices of the Royal Astronomical Society, 500, 1114,
  \dodoi{10.1093/mnras/staa3353}

\bibitem[{{Sprenger} {et~al.}(2021){Sprenger}, {Wucknitz}, {Main}, {Baker}, \&
  {Brisken}}]{swm+21}
{Sprenger}, T., {Wucknitz}, O., {Main}, R., {Baker}, D., \& {Brisken}, W. 2021,
  \mnras, 500, 1114, \dodoi{10.1093/mnras/staa3353}

\bibitem[{{Stinebring} {et~al.}(2001){Stinebring}, {McLaughlin}, {Cordes},
  {Becker}, {Goodman}, {Kramer}, {Sheckard}, \& {Smith}}]{smc+01}
{Stinebring}, D.~R., {McLaughlin}, M.~A., {Cordes}, J.~M., {et~al.} 2001,
  \apjl, 549, L97, \dodoi{10.1086/319133}

\bibitem[{van Kerkwijk \& van
  Lieshout(2022)}]{marten_h_van_kerkwijk_2022_7455536}
van Kerkwijk, M.~H., \& van Lieshout, R. 2022, mhvk/screens: v0.1, v0.1,
  Zenodo, \dodoi{10.5281/zenodo.7455536}

\bibitem[{{Walker} {et~al.}(2004){Walker}, {Melrose}, {Stinebring}, \&
  {Zhang}}]{wmsz04}
{Walker}, M.~A., {Melrose}, D.~B., {Stinebring}, D.~R., \& {Zhang}, C.~M. 2004,
  MNRAS, 354, 43, \dodoi{10.1111/j.1365-2966.2004.08159.x}

\bibitem[{{Zhu} {et~al.}(2023){Zhu}, {Baker}, {Pen}, {Stinebring}, \& {van
  Kerkwijk}}]{Zhu2023}
{Zhu}, H., {Baker}, D., {Pen}, U.-L., {Stinebring}, D.~R., \& {van Kerkwijk},
  M.~H. 2023, \apj, 950, 109, \dodoi{10.3847/1538-4357/accde0}

\end{thebibliography}
%\bibliographystyle{aasjournal}
%\input{B1737two-screen.bbl}

\end{document}